# The Effect of Investor-Driven Information Diffusion on Excess Comovement: Evidence from Retail and Institutional Investors in China and the United States

Fei Ren [a], Miao-Miao Yi [a], Zhang-Hangjian Chen [b], Xiang Gao[c,*]

[a]*School of Business, East China University of Science and Technology, Shanghai 200237, China*

[b]*School of Economics, Anhui University, Hefei 230601, China*

[c] *Research Center of Finance, Shanghai Business School, Shanghai 200235, China*

***Abstract:*** This study investigates how cross-stock information diffusion, driven by both retail and institutional investors, influences excess comovement in the Chinese retail-dominated market and the U.S. institution-dominated market. Using data from 4,533 Chinese stocks and 4,517 U.S. stocks from 2010–2022, we identify three key findings. First, the dominant investor group in each market significantly drives excess comovement. Specifically, in China, compared with institution-driven diffusion, retail-driven information diffusion has a notably stronger effect on excess comovement. In contrast, in the U.S., institution-driven diffusion is the primary driver of excess comovement, surpassing the influence of retail-driven diffusion. Second, we identify investors' trading behavior as the underlying mechanism through which information diffusion affects excess comovement. Third, we observe a lead-lag relationship: stocks with faster retail-driven information diffusion exhibit comovement that precedes those with slower diffusion. Based on this finding, we further demonstrate that the predictive power of information diffusion varies across markets. In China, retail-driven diffusion shows strong and persistent predictability for excess comovement, whereas in the U.S., institution-driven diffusion exhibits similarly robust predictive capacity.

## 1.Introduction

Excess comovement refers to the part of contemporaneous stock return correlations that cannot be explained by common fundamentals. Given its critical implications for risk management and portfolio allocation, identifying its potential drivers has become a focal point for researchers. The literature demonstrates that information channels between firms, such as common analyst coverage, corporate disclosure similarity, and media coverage, contribute to excess comovement (Israelsen, 2016; Box, 2018; Chen et al., 2021). These channels reflect the supply-side information provided by information producers and intermediaries. Recent studies have identified a demand-side information channel through investors' correlated searches for related stocks on platforms like Yahoo! Finance and Google. By searching for information on these related stocks, investors facilitate cross-stock information diffusion, thereby inducing excess comovement (Lee et al., 2015; Agarwal et al., 2017; Tan, 2022).

Motivated by the investor-centric perspective, this study shifts the focus from information demand to information consumption, examining how retail investors' interactions on stock forums contribute to excess comovement. Retail investors are increasingly relying on social media, particularly stock forums, for investment ideas[1]. After processing investment-related information, investors tend to share and discuss such information with one another through posts and replies (Chen et al., 2014). Given limited attention (Egeth and Kahneman, 1975; Hirshleifer and Teoh, 2003), they are active in limited subforums corresponding to stocks that share similar fundamentals. When investors participate in multiple subforums, their interactions transmit relevant information across these subforums, promoting information diffusion between related stocks and ultimately leading to excess comovement. Unlike supply-side or demand-side information, we capture investors' information consumption through posts and replies, which reflect their interpretations of information (Blankespoor et al., 2020). Furthermore, by tracking investors' interactions across subforums, we can directly observe the pathways of information diffusion between stocks and accurately measure retail-driven cross-stock information diffusion.

To examine whether the impact of retail investors on excess comovement persists after accounting for information diffusion driven by institutional investors, we conduct a joint analysis of both investor types. Researchers find that common institutional ownership—ownership by institutional investors who simultaneously hold significant stakes in at least two firms—promotes information diffusion between commonly owned stocks, causing their returns to move together (Cohen and Frazzini, 2008; Menzly and Ozbas, 2010; Anton and Polk, 2014; Gao et al., 2017; Cen et al., 2025). By holding significant stakes in multiple firms, common institutional investors aggregate information from corresponding stocks and use it to manage their portfolios. As informed traders, they tend to adjust their ownership of a stock simultaneously at key inflection points of stock prices, such as executive leadership transitions, technological breakthroughs, and regulatory shocks (Ying, 2024). This adjustment is particularly sensitive to the price movements of other commonly owned stocks, leading to stock prices fluctuating in the same direction and generating comovement. Retail-driven information diffusion originates from forum interactions, whereas institutional information diffusion is embedded within ownership networks. Our joint analysis of retail and institutional investors helps reveal how different investor types affect excess comovement by facilitating cross-stock information diffusion.

Market microstructure theory (Mann and O'Hara, 1996) suggests that the dominant investor group in a market exerts disproportionate influence on stock prices. The Chinese and U.S. stock markets, as the two largest global markets, exhibit diametrically opposed investor structures. The Chinese stock market is dominated by retail investors, who account for 84.5% of total trading volume. In contrast, the U.S. stock market is institutionally dominated, with institutional investors holding 68.3% of total market capitalization (Jones et al., 2024). This structural difference provides an ideal setting for examining how cross-stock information diffusion, driven by retail and institutional investors, impacts excess comovement under different market structures. Specifically, we investigate whether retail investors have a stronger effect on excess comovement than institutional investors in the Chinese market. Similarly, we explore whether in the U.S. market, institutional investors exert a more pronounced influence on excess comovement than retail investors do.

[1]"Social media is the most popular source of investment ideas for young investors, CNBC survey finds," CNBC, August 26, 2021; https://www.investor.org.cn/investor_interaction/questionnaire/tzzbg/201905/t20190509_365638.shtml.

Our dual samples consist of 4,533 Chinese stocks and 4,517 U.S. stocks from 2010 to 2022. Stock forum data for these stocks are sourced from Eastmoney in China and StockTwits in the U.S., while institutional holdings are obtained from quarterly disclosures. Following Cohen and Frazzini (2008), we use common institutional ownership as a proxy for institution-driven cross-stock information diffusion. For retail investors, we focus on co-investors who are active in two subforums within a specific period. We measure retail-driven information diffusion by using the number of their posts and replies, as well as the replies triggered by their posts and replies. Excess comovement is calculated as the pairwise Pearson correlations of stock-level residual returns from the Fama-French five-factor model (Fama and French, 2015). To account for other determinants of comovement, we include industry, location, price, size, return on assets, leverage, volatility, momentum, analyst coverage, and media coverage as control variables.

Our baseline analysis examines the effect of investor-driven cross-stock information diffusion on excess comovement. In China's retail-dominated market, compared with institutional diffusion, retail-driven information diffusion has a stronger impact on excess comovement. This finding aligns with the dominance of retail investors' trading volume in China. In contrast, U.S. results show that institution-driven diffusion is the primary driver of excess comovement, consistent with the predominant market capitalization share held by institutional investors. These divergent cross-market results arise from how dominant investor groups contribute to excess comovement. In China, retail investors' active information interactions on stock forums lead to similar trading behavior that synchronizes price movements, whereas in the U.S., common institutional ownership triggers excess comovement through simultaneous portfolio reallocation.

A major concern regarding our results is that the impact of information diffusion on excess comovement may be endogenously determined. Although our baseline specifications include time, industry and firm fixed effects to mitigate omitted variable bias, we rigorously address potential endogeneity through three identification strategies. First, to establish causal identification, we use a quasi-natural experiment based on the launches of the Eastmoney mobile application in China (November 2012) and the StockTwits mobile application in the U.S. (October 2010)—exogenous shocks that substantially increased forum interactions without directly influencing the stock market. We find that the impact of information diffusion on excess comovement becomes more pronounced after these events. Next, to control for systematic differences among our sample stocks, we perform the propensity score matching method and confirm that our results remain robust. Finally, we employ the instrumental variable approach to address potential sample selection bias. Our results indicate that the effect of information diffusion on excess comovement remains robust after addressing possible endogeneity. We further conduct a series of robustness checks, including substituting the measure of excess comovement, recalculating institution-driven information diffusion and control variables, incorporating high-dimensional fixed effects, excluding stock pairs within the same industry, removing observations with low forum interactions, and applying the Fama-Macbeth (FM) regression to reexamine our findings. Overall, all these tests support our baseline results, confirming that investor-driven information diffusion contributes to excess comovement.

We further explore the underlying mechanism through which information diffusion induces excess comovement. Following Kumar and Lee (2006), we measure the consistency of investors' trading behaviors using the Pearson correlation of the buy-sell imbalance (BSI) between stock pairs. We then examine whether investors' trading behavior can explain the impact of information diffusion on excess comovement. Our findings suggest that investors' trading behavior is a key mechanism in both markets. Consequently, our study provides empirical evidence that investors' trading behavior serves as a channel through which information diffusion influences excess comovement.

Our final analysis investigates the predictive power of information diffusion on excess comovement through two complementary approaches. First, we examine the lead-lag relationships between excess comovements in stock returns caused by varying speeds of information diffusion. Specifically, we classify stock pairs into quintiles based on information diffusion intensity and calculate the weighted average correlation coefficient for each group according to the magnitude of information diffusion. We find that the correlation coefficient for the group with the fastest information diffusion is significantly greater than that for the group with the slowest information diffusion. Based on these findings, we develop a lagged predictive model to assess the ability of information diffusion to forecast excess comovement. The results indicate that in China, retail-driven information diffusion exerts a strong and persistent predictive effect on excess comovement across multiple quarters, significantly surpassing the weak and transient predictive influence of institutional diffusion. In contrast, in the U.S., institutional diffusion demonstrates robust predictability across multiple quarters, whereas retail diffusion exhibits only transient predictive power confined to a single quarter. Collectively, these findings underscore how information diffusion influences excess comovement in markets with different investor dominances.

Our study contributes to several strands of research. First, we extend the excess comovement literature by developing an information consumption framework that complements prevailing supply-side explanations. Existing literature has predominantly focused on supply-side information channels to explain excess comovement, such as common analyst coverage (Israelsen, 2016), corporate disclosure similarity (Dyer et al., 2023), and media coverage (Chen et al., 2021). Motivated by Agarwal et al.'s (2017) finding that investors' correlated searches act as demand-side information channels, we demonstrate how investors' forum interactions facilitate cross-stock information diffusion and induce excess comovement from the perspective of information consumption. Unlike reports or searches, forum interactions capture information consumption through posts and replies, which reflect investors' interpretation of information. Furthermore, by tracing co-investors' interactions across subforums, we uncover the pathways of information diffusion between related stocks.

Second, our study enriches institution-centric explanations of excess comovement by distinguishing the mechanisms through which different types of investors influence excess comovement. Most studies on cross-stock information diffusion emphasize the role of institutional investors, showing that the information embedded in common institutional ownership significantly affects excess comovement (Cohen and Frazzini, 2008; Menzly and Ozbas, 2010; Anton and Polk, 2014; Gao et al., 2017; Ying, 2024). Moving beyond this institutional focus, we perform a comparative analysis that simultaneously examines the impact of retail-driven and institution-driven cross-stock information diffusion on excess comovement. We show that retail investors' forum-based information diffusion and institutional investors' ownership-linked information diffusion exert different effects on excess comovement.

Third, this study innovatively investigates investor-driven cross-stock information diffusion in two distinct markets. Previous studies have analyzed the investor composition, market segments, and other aspects of the two largest stock markets in the world, i.e., China and the U.S., revealing significant similarities and differences between them (Boehmer et al., 2021; Jones et al., 2024; Tan et al., 2024). Despite representing developing and mature markets, respectively, empirical research examining excess comovement has largely focused on single-market contexts. We address this gap by exploring how information diffusion influences excess comovement in both markets. Our findings indicate that, in the Chinese retail-dominated market, compared with institutional investors, retail-driven information diffusion has a stronger impact on excess comovement. In contrast, in the U.S. market, institution-driven information diffusion has greater explanatory power for excess comovement than retail-driven diffusion.

Finally, our mechanism analysis demonstrates that investors' trading behavior serves as the underlying channel through which information diffusion influences excess comovement, thereby bridging two strands of explanations for excess comovement. On the one hand, prior research indicates that investors' trading behavior is largely driven by information (Leung et al., 2016; Jiang et al., 2019). On the other hand, studies have established that trading behavior is a key driver of excess comovement (Barberis and Shleifer, 2003; Coval and Stafford, 2007). Consequently, by investigating whether trading behavior mediates the relationship between information diffusion and excess comovement, our study synthesizes the literature on excess comovement into a unified framework, integrating both information-channel and trading-behavior perspectives.

Related work by Jiang et al. (2019) shows that investors' communication on stock forums is associated with return comovement. Our study extends Jiang et al.'s (2019) work through methodological innovations in measuring cross-stock information diffusion. While they proxy communication by counting posts mentioning other stocks' tickers or names within a specific forum, we track co-investors' activities across related forums, thereby capturing three aspects that their approach may overlook. First, by analyzing co-investors' posts and replies as well as the replies they elicit, we identify information diffusion through abbreviations, concepts, and contextual references—phenomena particularly prevalent in the Chinese market. Second, whereas they focus on the top five attention-grabbing stocks, we comprehensively map all forum discussions involving co-investors. Finally, by restricting our analysis to investors active on related forums, we exclude noise from economically unrelated mentions, an inherent limitation of keyword-centric approaches. These enhancements enable us to identify more extensive and precise patterns of information diffusion.

The remainder of this paper is organized as follows. Section 2 outlines the institutional background, literature review and research hypotheses. Section 3 details the data for the two markets and the construction of variables. Section 4 presents and discusses the empirical results. Section 5 concludes.

## 2. Institutional Background, literature review and hypothesis development

### *2.1 Institutional Background*

#### *2.1.1 Market structures in the Chinese and U.S. stock markets*

The Chinese stock market, established in 1990, has developed a hierarchical market structure over three decades of rapid growth. The Shanghai and Shenzhen mainboards, similar to the NYSE in the U.S. market, primarily serve large, well-established enterprises; the small and medium enterprise (SME) board is specifically designed to support small and emerging enterprises; and the growth enterprise market (GEM) board, launched in 2010 as "China's NASDAQ," focuses on high-growth technology enterprises. As the world's second-largest stock market by total capitalization after the U.S., Chinese A-share market reached over 80 trillion yuan in market capitalization by the end of 2023. Retail investors account for approximately 80% of the daily trading volume but hold only about 20% of the market capitalization. Conversely, institutional investors own approximately 20% of the market capitalization but contribute less than 20% to overall trading activity (Jones et al., 2024).

In contrast to China, the U.S. stock market features a markedly different investor structure, being predominantly shaped by institutional investors. According to Boehmer et al. (2021), retail investors account for about 16% of daily trading activity and hold a relatively small share of total market capitalization. As disclosed in 13F filings, institutional investors possess a substantial ownership of 67% market capitalization. Thus, in the U.S., institutional investors, rather than retail investors, are the primary trading and holding entities. Notably, the advent of fractional trading and the rise of zero-commission trading platforms, such as Robinhood, have significantly enhanced the influence of retail investors in recent years (Da et al., 2025). During the 2021 meme stock frenzy, retail investors accounted for as much as 30% of trade activity (Barber et al., 2022). Despite this significant increase in retail investor activity, institutional investors continue to dominate the U.S. market.

The differences in market structures between China and the United States provide an ideal setting for examining how investor-driven cross-stock information diffusion triggers excess comovement. In China, retail investors exchange information on social media platforms, which may increase the return comovement of related stocks, whereas institutional investors, subject to regulatory constraints, may exert less influence than retail investors (Allen et al., 2005). In the U.S., institutional holdings may strengthen return correlations across stocks through their similar strategies. Although retail investors tend to be more active around specific events, their relatively small holdings and trading volumes likely limit their effect on excess comovement. Existing research has predominantly focused on comovement driven by institutional investors in developed markets (Cohen and Frazzini, 2008; Menzly and Ozbas, 2010; Anton and Polk, 2014; Gao et al., 2017), leaving the phenomenon of retail-driven comovement in emerging markets underexplored. This study utilizes a dual-sample comparison to deepen understanding and offer significant insights for regulatory practices.

*2.1.2 Retail investors on social media: Guba in China and StockTwits in the U.S.*

China has an internet user base exceeding one billion and a highly dynamic social media landscape (Li et al., 2024). Among various social media platforms, online stock forums are the most popular channel for retail investors to share information. Guba.EastMoney.com, hereafter referred to as Guba, is China's largest and most influential stock forum. Since its launch in 2005, Guba has attracted over 120 million registered users who generate more than three million posts per day. The platform offers specialized subforums for individual stocks, enabling investors to engage in real-time discussions regarding these stocks by posting and replying. These information interactions cover topics including financial performance, corporate governance, and policy interpretation. Its institutional recognition is evidenced by a 2018 survey by the CSI Small and Medium Investor Service Center[2], which reports that 80% of retail investors consider Guba their primary channel for both gathering and disseminating stock-related information.

Additionally, existing studies characterize posting activity on stock forums—particularly on Guba, a representative platform—as a form of retail investor behavior (Huang et al., 2016; Ang et al., 2021; Li and Zhang, 2022). Therefore, it is reasonable to measure the information dissemination behavior of retail investors using their posts and replies on stock forums. On

[2]See https://www.investor.org.cn/investor_interaction/questionnaire/tzzbg/201905/t20190509_365638.shtml for more information.

September 13, 2021, the delayed redemption crisis of Evergrande Wealth's financial products was revealed, triggering intense discussions in stock forums related to Evergrande's suppliers, real-estate partners, and creditor banks. As these communications spread rapidly, the stock prices of the affected firms fell sharply, indicative of market panic exacerbated by retail-driven information diffusion. The real-time and high-volume information flows on Guba position it as a vital observatory for analyzing excess comovement driven by retail investors in the Chinese market.

In the U.S. market, StockTwits.com serves as the primary platform for studying how retail investors promote cross-stock information diffusion. Launched in 2008, StockTwits is a prominent finance-focused social media platform similar to Twitter but specifically designed for investors and traders. Since its inception, StockTwits has attracted significant attention from major news media outlets such as The New York Times and CNNMoney.com. By 2020, users generated over 6.5 million posts per month and numbered more than 2 million unique accounts, including about 400,000 active posters (Cookson et al., 2022). When users want to post content about a company on the StockTwits website, they tag the company's stock code using the "$" symbol. For example, the post "$GOOG and $MSFT, you should buy!" explicitly signals positive recommendations for Alphabet and Microsoft shares. Through the use of stock tags, the companies referenced in the posts can be identified unambiguously. By the end of 2022, cumulative posts exceeded 300 million, making StockTwits one of the largest professional financial social media platforms globally.

Following the GameStop mania in 2021, researchers have shown growing interest in the impact of retail investors on the U.S. stock market. Although StockTwits cannot fully represent retail investors across the broader market, its external validity is well-supported by numerous studies, positioning it as the preferred platform for analyzing retail investors' behavior (Giannini et al., 2019; Cookson and Niessner, 2020; Cookson et al., 2024; Hirshleifer et al., 2025; Vamossy and Skog, 2025). Therefore, StockTwits offers a highly credible observational setting for analyzing the information diffusion among retail investors in the United States.

### *2.2 Literature review and hypothesis development*

#### *2.2.1 Information diffusion and excess comovement*

Roll's (1988) pioneering research reveals that a substantial portion of stock return volatility remains unexplained by market factors or publicly available information. This phenomenon can be attributed to informed traders who utilize private information to make investment decisions, thereby exerting distinct influences on stock prices. Based on this finding, Barberis et al. (2005) propose the information diffusion hypothesis to explain excess comovement. According to this hypothesis, information is incorporated into stock prices at different rates due to market frictions, leading to greater comovement among stocks that absorb information at similar speeds. For example, stocks held by investors with information advantages tend to incorporate and reflect new information more rapidly. When relevant information is released, these stocks exhibit contemporaneous return correlations, whereas other stocks experience delayed price adjustments due to slower information diffusion. Subsequent studies have further explored comovement through the lens of information channels. Given the difficulty of accessing investors' private information sets, researchers have instead focused on supply-side information provided by information producers.

From the supply-side perspective, the literature identifies three dominant information channels: common analyst coverage, corporate disclosure similarity, and media coverage. Analysts, as key information intermediaries, play a crucial role in influencing stock prices by collecting, analyzing, and disseminating relevant information. Studies show that when analysts simultaneously report on multiple stocks, their earnings forecasts often highlight common factors among related stocks based on shared models, methods, or data (Muslu et al., 2014; Israelsen, 2016), thereby leading to return correlations among the covered stocks. Corporate disclosure similarity provides another avenue for information diffusion (Box, 2018). Dyer et al. (2023) find that when companies disclose the same risk factors in their financial reports, this similarity draws investors' attention to related information, thereby strengthening cross-stock information diffusion and generating excess comovement. News coverage involving multiple stocks reflects journalists' interpretations of economic linkages between firms. Chen et al. (2021) point out that news discussing two stocks may reflect the complex economic ties between the firms, such as strategic cooperation, industry competition, and financing relationships, which promote information diffusion between stocks and effectively explain excess comovement (Chen and Wang, 2024; Ge et al., 2025).

Recent technological advancements have enabled researchers to track investors' digital footprints on online platforms, thereby shifting their focus to the demand side of information. By analyzing investors' proactive information–searching behavior for multiple stocks on Yahoo! Finance, Leung et al. (2016) find that stocks searched together by investors exhibit significant excess comovement. Their further analysis indicates that correlated searches reflect investors' collective attention to these stocks, serving as a proxy for the extent of information diffusion across stocks (Agarwal et al., 2017). Tan et al. (2022) apply this method to the Google search platform, confirming that investors' collective search behavior significantly enhances the return correlations of the co-searched stocks. Motivated by this investor-centric perspective, this study further shifts the focus from information demand to consumption, exploring whether retail investors' information interactions in stock forums promote cross-stock information diffusion and generate excess comovement. Unlike correlated searches, forum interactions reflect investors' interpretations of information, potentially becoming a more direct channel for information diffusion between stocks.

Our study of the impact of investors' forum interactions on excess comovement is also motivated by the word-of-mouth effect, which refers to the process where individuals actively share information, opinions, and experiences via face-to-face communication. Prior research demonstrate that a similar word-of-mouth effect exists within investor groups. Hong et al.

(2005) report that fund managers in the same city frequently exchange investment ideas through local meetings, and this sharing of information directly influences their trading decisions on specific stocks. Ivković and Weisbenner (2007) further validate this effect among retail investors, revealing that household investment decisions are significantly influenced by their neighbors—each 10% increase in neighbors' holdings of a particular stock leads to an approximately 2% increase in households' purchases of that stock. In the internet age, stock forums transcend physical and geographical limitations, enabling investors to more conveniently share and exchange stock-related information, such as financial performance, corporate governance, and policy interpretation. Consequently, these platforms serve as an online embodiment of word-of-mouth communication.

Based on the findings of these studies, we investigate how retail investors' information interactions in stock forums influence excess comovement. Given limited attention (Hirshleifer and Teoh, 2003), investors tend to focus on a limited set of stocks with similar fundamentals, such as size, industry and location. This focus leads them to be active in the corresponding subforums. We focus on co-investors, who are active in two subforums within a specific period, as they serve as intermediaries for information diffusion between related stocks. Due to the similarities between stocks, posts or replies in one subforum may contain relevant information about the other stock, such as important business plans, merger announcements, and industry policies. When co-investors post or reply in both subforums, this information may be transmitted from one stock to the other. Additionally, responses triggered by these posts or replies from other investors can further facilitate information diffusion between the two stocks. These interactions contain investors' interpretations of information (Blankespoor et al., 2020), thereby reflecting the diffusion of that portion of information consumed by investors. Therefore, we use the number of co-investors' posts and replies and the replies triggered by their posts and replies to measure the strength of cross-stock information diffusion. The greater the number of posts and replies that co-investors trigger, the stronger the information diffusion between the two stocks and the higher the degree of excess comovement it induces. Therefore, we propose the following hypothesis:

**H1:** Retail investors' information interactions in stock forums promote cross-stock information diffusion, thereby contributing to excess comovement.

Retail investors promote cross-stock information diffusion via forum interactions, whereas institutional investors facilitate such diffusion through ownership network. The literature highlights that common institutional ownership establishes interfirm information channels, which contributes to excess comovement (Cohen and Frazzini, 2008; Menzly and Ozbas, 2010; Anton and Polk, 2014). By holding substantial equity stakes in multiple stocks, common institutional investors facilitate information diffusion between the commonly held stocks through portfolio reallocation, thereby inducing return correlations among these stocks (Cohen and Frazzini, 2008; Anton and Polk, 2014). Specifically, common institutional investors acquire relevant information from the stocks they hold for portfolio management.

Leveraging their privileged access to information, they adjust their ownership of a stock simultaneously in response to good or bad corporate news (Ying, 2024). This adjustment in institutional ownership is particularly sensitive to the prices of other stocks held by common institutional investors, leading to comovements in stock prices. Compared with retail investors, information diffusion driven by institutional investors is embedded within a long-term, stable ownership network, which may make their impact on excess comovement more persistent. Therefore, we propose the following hypothesis:

**H2:** Common institutional ownership facilitates cross-stock information diffusion, which amplifies excess comovement.

Existing research highlights substantial differences in investor structures between the Chinese and U.S. stock markets (Jones et al., 2024; Tan et al., 2025), providing a natural experimental setting to examine the distinct effects of two types of investors on excess comovement. In China, retail investors play a dominant role in the market, accounting for approximately 80% of the daily trading volume. Moreover, their information interactions on Guba generate more than 3 million posts each day (Li et al., 2024). This highly active trading and communication ecosystem promotes retail-driven cross-stock information diffusion, enabling rapid responses to market events and inducing excess comovement. In contrast, Chinese institutional investors face stringent regulations that impose restrictions on major shareholder sell-offs and short-term trading, potentially limiting their influence on excess comovement. In the U.S. market, institutional investors control 67% of market capitalization, and may exert a relatively stronger impact on excess comovement than retail investors do. Although zero-commission platforms such as Robinhood amplify retail investors' influence during meme stock events (Barber et al., 2022), their fragmented shareholding may hinder the formation of sustained cross-stock information diffusion, thereby diminishing their impact on excess comovement. Based on these studies and the distinct market structures observed in the two countries, we propose the following hypotheses:

**H3a:** In the Chinese market, compared to institutional investors, retail-driven cross-stock information diffusion has a more substantial impact on excess comovement.

**H3b:** In the U.S. market, institution-driven cross-stock information diffusion demonstrates stronger explanatory power for excess comovement than that driven by retail investors.

*2.2.2 Mechanism through which information diffusion affects excess comovement*

A foundational premise in financial economics posits that information events, such as firm announcements, earnings reports, and other news releases, must be mediated through investors' trading activities to become incorporated into stock prices (Grossman and Stiglitz, 1980). This mediation process is particularly critical for explaining how information diffusion leads to excess comovement. Leung et al. (2016) posit that investors' trading behavior may serve as a critical mechanism through which information diffusion triggers excess comovement, but leave this mechanism untested and conclude by calling for future research to examine this proposition empirically.

On the one hand, information diffusion has a significant impact on investors' trading behavior. For retail investors, social media greatly amplifies the effect of information diffusion on their trading decisions. Cookson et al. (2022) find that discussions about related stocks on platforms such as StockTwits prompt many small investors to buy or sell the same stocks simultaneously. Similarly, Hirshleifer et al. (2025) demonstrate that when investors are exposed to identical information sources, such as industry news or popular forum posts, their buying and selling trading decisions become surprisingly similar. This phenomenon is particularly pronounced during major news events. For instance, when companies release earnings reports or governments announce policy changes, investors often follow trending forum discussions to decide what and when to trade (Ammann and Schaub, 2020). For institutional investors, common ownership networks serve as a channel for obtaining valuable information from the invested companies and accordingly adjust their ownership across related stocks (Ying, 2024).

On the other hand, investors' trading behavior directly influences excess comovement. Research shows that trading behavior of common institutional investors can lead to excess comovement. Anton and Polk (2014) emphasize that the strategic portfolio reallocation of common institutional investors, such as sector rotation and reactions to regulatory shocks, can result in excess comovement through order flow resonance. Geraci et al. (2023) further show that when a large number of common institutional investors engage in short selling of related stocks, the coherence of order flows can significantly increase the return correlation of these stocks. With respect to retail investors, Umar et al. (2021) find that during the GameStop frenzy in 2021, the coordinated buying behavior of investors on Reddit not only caused a substantial increase in the target stock price but also induced return correlations among related stocks. These two strands of literature support our investigation into investors' trading behavior as the mechanism by which information diffusion leads to excess comovement. Therefore, we propose the following hypothesis:

**H4:** Investors' trading behavior serves as the mechanism through which information diffusion induces excess comovement.

## 3. Data, Sample and Variables

### *3.1 Data and Sample*

This study examines stocks listed on major exchanges in China and the United States to ensure comparability across markets. In the Chinese market, the sample includes all common A-share stocks listed on the Shanghai Stock Exchange (SSE) and Shenzhen Stock Exchange (SZSE), covering diverse market segments: the SSE and SZSE main boards for large-cap firms, the SME board for small-to-medium enterprises, and the GEM board for high-growth technology companies. In the U.S. market, the sample includes stocks listed on the NASDAQ, NYSE, and AMEX, representing corresponding market tiers: the NYSE for established firms, the NASDAQ for technology stocks, and the AMEX for emerging companies.

Our research period spans from 2010 to 2022, covering multiple market cycles to enhance statistical power. To ensure data quality and representativeness, we apply two key sample filters: stocks with forum message volume below 10% of the average message volume across all forums are excluded to ensure sufficient retail investors' interactions, and newly listed firms undergoing initial public offerings (IPOs) within the current quarter are excluded to reduce the impact of anomalous price volatility. After applying these filters, the final dataset consists of 4,533 Chinese stocks and 4,517 U.S. stocks, generating approximately 180 million stock-pair observations for China and 140 million for the U.S. across 52 quarterly periods.

Our data are compiled from multiple sources to ensure comprehensive coverage. Retail investors' interactions are obtained from leading specialized stock forums: Eastmoney Guba for Chinese stocks and StockTwits for U.S. stocks. We collect full historical records of posts, replies, user IDs, and timestamps via customized web scrapers, yielding more than 600 million messages. Institutional ownership data are sourced from CSMAR (China) and Thomson Reuters 13F (U.S.). Market transaction data—including daily capital inflows and outflows, tick-by-tick transactions, daily closing prices, and Fama-French factors—are obtained from Wind (China) and TAQ/CRSP (U.S.). Fundamental data on firm characteristics, such as market capitalization, industry classification, headquarters location, return on assets, stock return, and leverage, are sourced from CSMAR (China) and Compustat (U.S.). Additionally, news coverage data are obtained from CSMAR (China) and RavenPack Analytics (U.S.), while analyst coverage data are derived from CSMAR (China) and IBES(U.S.), helping to account for the influence of alternative information channels.

Notably, there is substantial heterogeneity in data frequencies: institutional ownership and firm characteristics are naturally reported on a quarterly basis, whereas retail investors' interactions, news coverage, analyst reports, stock prices, and transactions are recorded on a daily or intraday basis. To align these data temporally with institutional reporting cycles and reduce the risk of asynchronous measurement bias, all high-frequency variables are aggregated to a quarterly level using appropriate methodological approaches. Specifically, summation is applied to forum posts and counts of news or analyst reports, while correlation measures are employed to capture excess comovement and the consistency of investors' trading behavior.

### *3.2 Variables*

#### *3.2.1 Measure of excess comovement*

In this study, we define excess comovement as the pairwise correlation of residual returns between two stocks after controlling for fundamental and systematic risk factors. In line with the literature on excess comovement (Israelsen, 2016; Li et al., 2019; Chen et al., 2021), we employ the Fama-French five-factor (FF5) model (Fama and French, 2015) to isolate systematic components of returns. For stock $i$ on trading day $d$, the excess return residual $e_{id}$ is computed as follows:

$$r_{id} - r_{fm} = \alpha_0 + \alpha_1 MKT_d + \alpha_2 SMB_d + \alpha_3 HML_d + \alpha_4 RMW_d + \alpha_5 CMA_d + e_{id}, \quad (1)$$

where $r_{id}$ denotes the return of stock $i$ on day $d$; $r_{fm}$ represents the risk-free rate; and $MKT_d$, $SMB_d$, $HML_d$, $RMW_d$, and $CMA_d$ correspond to market, size, value, profitability, and investment factors, respectively.

Based on the pairwise methodology developed by Kallberg and Pasquariello (2008), the excess comovement for stock pair $i$ and $j$ in quarter $t$ is quantified by calculating the Pearson correlation coefficient of their daily returns:

$$Cor_{ij,t}^{FF5} = \frac{\sum_{d=1}^{D_t} e_{id}\, e_{jd}}{\sqrt{\left(\sum_{d=1}^{D_t} e_{id}^2\right)\left(\sum_{d=1}^{D_t} e_{jd}^2\right)}}, \tag{2}$$

where $D_t$ represents the number of trading days in quarter $t$, and $e_{id}$ and $e_{jd}$ denote the daily residual returns from the Fama-French five-factor (FF5) model for stocks $i$ and $j$ on day $d$, respectively. $Cor_{ij,t}^{FF5}$ represents the excess comovement between stock pair $i$ and $j$ during quarter $t$. To ensure robustness, we also compute excess comovement using both the Fama-French three-factor (FF3) model (Fama and French, 1993) and the capital asset pricing model (CAPM) (Sharpe, 1964) as part of our robustness tests.

*3.2.2 Measure of retail-driven information diffusion*

Retail-driven information diffusion is measured through investors' interactions on stock forums. In this study, we focus on co-investors who serve as conduits for the transmission of information between related stocks. Their posts and replies facilitate the diffusion of relevant information across these forums, leading investors to possess similar information about the related stocks. This similarity in information may promote correlated trading behavior, thereby contributing to excess comovement. Accordingly, we aggregate the number of co-investors' posts and replies, as well as the subsequent replies triggered by those posts and replies, to measure the intensity of retail-driven information diffusion across both forums. We first identify co-investors as users who post or reply on the forums corresponding to both stocks $i$ and $j$ during quarter $t$. Then we calculate the information flow metric $Flow_{ij,t}$ as follows:

$$Flow_{ij,t} = \sum_{n=1}^{N_{ij}} (Po_{ni,t} + Po_{nj,t} + Re_{ni,t} + Re_{nj,t}), \tag{3}$$

where $N_{ij}$ represents the number of co-investors who are active on both forums during quarter $t$; $Po_{ni,t}$ and $Po_{nj,t}$ denote the number of posts made by co-investor $n$ on forums $i$ and $j$ during quarter $t$, respectively; and $Re_{ni,t}$ and $Re_{nj,t}$ represent the number of replies triggered by co-investor $n$'s posts on forums $i$ and $j$ during quarter $t$, respectively. To address concerns regarding inflated measures due to low-engagement content, we exclude posts with fewer than 3 replies in the robustness checks.

*3.2.3 Measure of institution-driven information diffusion*

Institution-driven information diffusion is captured through common ownership networks. By holding stakes in multiple firms, common institutional investors acquire private information through active monitoring (Gao et al., 2017; Lewellen, 2021). They strategically disseminate firm-specific information to other commonly owned firms to coordinate investment timing. Meanwhile, as informed traders, they tend to adjust their holdings of stocks simultaneously at key inflection points of stock prices. These adjustments transmit information across ownership networks, triggering excess comovement (Anton and Polk, 2014; Geraci et al., 2023). Following Cohen and Frazzini (2008), we measure institution-driven information diffusion using a common ownership indicator $Inst_{ij,t}$, defined as the sum of ownership proportions across all institutions that hold both stocks $i$ and $j$ during quarter $t$:

$$Inst_{ij,t} = \sum_{k=1}^{K_{ij}} (w_{ki,t} + w_{kj,t}), \tag{4}$$

where $K_{ij}$ denotes the number of institutions that hold both stocks $i$ and $j$ during quarter $t$, $w_{ki,t}$ represents institution $k$'s ownership proportion of stock $i$ (calculated as shares held divided by total outstanding shares, expressed as a percentage), and $w_{kj,t}$ denotes its ownership proportion in stock $j$. To enhance the robustness of our analysis, we employ alternative ownership thresholds to validate our findings.

*3.2.4 Measure of investors' trading behavior*

Investors' trading behavior is measured through a buy-sell imbalance (BSI) framework, which quantifies the synchronization of buy and sell decisions across stock pairs. Kumar and Lee (2006) develop the BSI indicator to capture investors' trading activities within a group of stocks and compute the correlation of the BSI for stock pairs as a measure of correlated trading. They find that systematic trading behavior among retail investors is associated with stock return comovement. Leung et al. (2016) propose that investors' trading behavior driven by information can lead to excess comovement among stocks. Jiang et al. (2019) show that stocks that are collectively discussed by investors on stock forums are traded more actively than those that are not. Therefore, it is reasonable to hypothesize that information diffusion promotes investors' trading behavior and contributes to excess comovement. To quantify this behavior, we adopt the methodology established by Kumar and Lee (2006) to calculate the daily BSI for stock $i$ and investor type $x$ (retail or institutional) as follows:

$$BSI_{id}^{x} = \frac{VB_{id}^{x} - VS_{id}^{x}}{VB_{id}^{x} + VS_{id}^{x}}, \tag{5}$$

where $VB_{id}^{x}$ and $VS_{id}^{x}$ represent the total daily dollar value (for the U.S. market) and yuan value (for the Chinese market) of buy and sell transactions for investor type $x$, respectively. The investor classification is market-specific to ensure methodological rigor. For the Chinese market, trades are classified as retail if the trade value is less than 40,000 yuan; all other trades are categorized as institutional, consistent with the classification guidelines provided by WIND. For the U.S. market, retail trades are identified using Boehmer et al.'s (2021) algorithm, and institutional trades are defined as transactions valued at 20,000 dollars (Lee and Radhakrishna, 2000).

We then construct the Pearson correlation coefficient index $COR_{BSI,ij,t}^{x}$ to measure the consistency of investors' trading behavior for stock pair $i$ and $j$ in quarter $t$, which is calculated as the correlation between $BSI_{id}^{x}$ and $BSI_{jd}^{x}$ for the corresponding stock pair:

$$COR_{BSI,ij,t}^{x} = \text{pearson}\left(BSI_{id}^{x}, BSI_{jd}^{x}\right), \tag{6}$$

where $BSI_{id}^{x}$ represents the daily BSI index for stock $i$ with respect to investor type $x$ (retail or institutional) on day $d$, $BSI_{jd}^{x}$ denotes the daily BSI index for stock $j$ with respect to investor type $x$ (retail or institutional) on day $d$, and $COR_{BSI,ij,t}^{x}$ denotes the Pearson correlation coefficient of the BSI index between stock pair $i$ and $j$ in quarter $t$.

*3.2.5 Measure of control variables*

Based on style investing theory (Barberis and Shleifer, 2003), investors categorize stocks into distinct style groups based on shared fundamental characteristics such as price, size, value, and profitability. Stocks within the same group face similar exposures to economic shocks and industry trends. As a result, when market-wide signals arise, investors tend to react collectively at the style level rather than the firm level, amplifying correlated trading and inducing excess comovement within these groups (Wahal and Yavuz, 2013; Kaul et al., 2016). To eliminate the possibility that the observed excess comovement is driven by similar fundamentals rather than information diffusion, we follow the established literature and include controls for fundamental similarity. Specifically, we include the variables *Sameind* (a dummy variable equal to 1 if stocks $i$ and $j$ belong to the same industry classification) and *Samelocation* (a dummy variable equal to 1 if the headquarters of stocks $i$ and $j$ are located in the same city). Furthermore, we construct several additional variables: *Similarprice*, *Similarsize*, *Similarroa*, *Similarlever*, *Similarvol*, and *Similarmom*. Each of these variables takes a value of 1 if stocks $i$ and $j$ share the same decile for price, market value of equity, return on assets, leverage ratio, volatility and annual stock return during the quarter (Dyer et al., 2023). To account for the potential influence of alternative information channels, we include controls for *News* (the number of news stories involving stocks $i$ and $j$) and *Analyst* (the number of analysts covering stocks $i$ and $j$). All control variables are measured quarterly at the stock-pair level. This design ensures that the observed excess comovement reflects investor-driven information diffusion rather than spurious correlations arising from shared fundamentals or external information channels. Appendix A specifies the details of variable definitions.

*3.3 Descriptive statistics*

Table 1 illustrates the divergent developmental trajectories of retail investors' activities on leading stock forums in China and the U.S. from 2010 to 2022. In China's rapidly evolving market, investor engagement on the Eastmoney platform has surged significantly, with a 220% increase in the number of actively discussed stocks, rising from 1,416 in 2010 to 4,533 by 2022. This substantial expansion in coverage amplifies retail-driven information diffusion tenfold, as evidenced by the growth in unique stock pairs, which increased from 986,820 to 10.26 million over the same period. While initial forum engagement was low, averaging only 1.94 posts and replies per pair in 2010, retail participation experienced a substantial increase after 2015. The number of co-investors per pair rose sevenfold, increasing from 7.38 in 2015 to 47.15 by 2021, and the number of posts and replies per pair peaked at 719.55 in 2019. This growth illustrates robust cross-stock information diffusion driven by retail investors, although activity was moderated post-2020 due to compounded challenges arising from the COVID-19 pandemic and regulatory tightening in China.

<Insert Table 1 Here>

In contrast, the U.S. stock market, which is characterized by more mature investor behavior on the StockTwits platform, shows steady retail participation. Despite a high baseline activity level of 33.36 posts and replies per pair in 2010, the growth rate remained relatively modest prior to 2020, with an annual growth rate of less than 15%. Structural inflection occurred during 2020–2021, when the GameStop event triggered a sharp acceleration in retail investor engagement. Specifically, the number of co-investors per pair experienced a significant year-on-year increase of 34.6%, rising from 52.28 in 2020 to 70.41 in 2021. Concurrently, the volume of posts and replies per pair increased by 7%, increasing from 605.17 to 647.33. The rise of commission-free platforms such as Robinhood further amplified retail-driven information diffusion, with engagement levels remaining near the peak values observed in 2021 (647.33 posts and replies per pair), indicating the ongoing penetration of retail influence in the U.S. market.

Table 2 presents descriptive statistics for key variables across the Chinese and U.S. markets. To enable a meaningful comparison of retail- and institution-driven information diffusion effects on excess comovement within each market, we standardize *Flow* and *Inst* as *Z_Flow* and *Z_Inst*, respectively, for regression analysis.

<Insert Table 2 Here>

These results highlight fundamental differences between the two markets. While U.S. stock pairs show a higher mean *Flow* (172.62) compared to China (106.94), its distribution is disproportionately concentrated, as indicated by a standard deviation of 408.702 and a maximum value of 622,955—both significantly higher than the 75th percentile value of 486. This pronounced right skew in the *Flow* distribution suggests that retail-driven information diffusion in the U.S. is predominantly influenced by outlier pairs, whereas China's tighter distribution reflects more consistent forum interactions.

Institution-driven information diffusion shows even more pronounced contrasts. In the U.S. market, stock pairs demonstrate significant common ownership, with an average of 48.68% and a median of 38.16%. In contrast, in China, stock pairs exhibit low institutional overlap, characterized by a median of 0% and a 75th percentile value of merely 0.006%. Standardized metrics further confirm this structural difference, with U.S. pairs showing a wider dispersion in *Z_Inst* values, ranging from −1.06 to 30.30, compared to China's range of −0.10 to 9.35. Excess comovement ($Cor^{FF5}$) further differentiates the markets. U.S. stock pairs show average correlations of 0.18 and a median of 0.20, which are significantly higher than China's mean of 0.01 and median of 0.00. Additionally, control variables—including media coverage (*News*), analyst coverage (*Analyst*), and fundamental characteristics such as *Similarsize* and *Similarprice*—show distributions that align with established literature on excess comovement.

*3.4 Empirical models*

To examine the respective and joint impact of retail- and institution-driven information diffusion on excess comovement, we construct the following baseline regression models, which correspond to Hypotheses 1 through 3:

$$Cor_{ij,t}^{FF5} = \beta_0 + \beta_1 Flow_{ij,t} + \beta_2 Control_{ij,t} + \varepsilon_{i,t} + \varepsilon_{j,t} + \epsilon_{ij,t}, \quad (7)$$

$$Cor_{ij,t}^{FF5} = \gamma_0 + \gamma_1 Inst_{ij,t} + \gamma_2 Control_{ij,t} + \varepsilon_{i,t} + \varepsilon_{j,t} + \epsilon_{ij,t}, \quad (8)$$

$$Cor_{ij,t}^{FF5} = \delta_0 + \delta_1 Flow_{ij,t} + \delta_2 Inst_{ij,t} + \delta_3 Control_{ij,t} + \varepsilon_{i,t} + \varepsilon_{j,t} + \epsilon_{ij,t}, \quad (9)$$

where the subscript $ij, t$ denotes stock pair $i$ and $j$ in quarter $t$ and $Cor_{ij,t}^{FF5}$ is the pairwise correlation of daily Fama-French five-factor residual returns for stock pair $i$ and $j$ in quarter $t$, representing excess comovement. $Flow_{ij,t}$ and $Inst_{ij,t}$ are the key independent variables capturing retail- and institution-driven information diffusion, respectively. $Control_{ij,t}$ denotes a set of control variables, including $News_{ij,t}$, $Analyst_{ij,t}$, $Similarroa_{ij,t}$, $Similarlever_{ij,t}$, $Similarsize_{ij,t}$, $Similarprice_{ij,t}$, $Similarmom_{ij,t}$, $Similarvol_{ij,t}$, $Samelocation_{ij,t}$, and $Sameind_{ij,t}$. Additionally, we include stock $i \times$ quarter $t$ and stock $j \times$ quarter $t$ fixed effects to absorb time-varying unobservable factors that may affect all pairs containing stock $i$ or $j$.

Equation (7) tests Hypothesis 1, which examines the effect of retail-driven information diffusion (*Flow*) on excess comovement. Hypothesis 2 is tested by estimating Equation (8), where we analyze the effect of institution-driven information diffusion (*Inst*) on excess comovement. Finally, Hypothesis 3 is tested by estimating Equation (9), which investigates the joint effect of retail- and institution-driven information diffusion on excess comovement.

## 4. Empirical Results

*4.1 Impact of investor-driven information diffusion on excess comovement*

We begin our empirical analysis by examining Hypothesis 1 (H1), which posits that retail-driven cross-stock information diffusion contributes to excess comovement. Columns (1) and (4) of Table 3 present the results from estimating Equation (7) for the Chinese and U.S. markets, respectively. As shown in the table, the coefficient on *Flow* is significantly positive in both markets (China: $\beta = 3.598$, $p < 0.01$; U.S.: $\beta = 0.010$, $p < 0.01$), providing strong support for Hypothesis 1. Additionally, we observe that the control variables for alternative information channels, including shared news coverage (*News*) and analyst coverage (*Analyst*), exhibit significantly positive coefficients. These results align with prior literature (Israelsen, 2016; Chen et al., 2021), reinforcing the conclusion that information linkages strengthen return correlations. Furthermore, the control variables capturing fundamental similarities, such as *Sameind*, *Similarroa*, *Similarsize*, and other relevant characteristics, are all positively significant, consistent with style-investing theory (Barberis and Shleifer, 2003) and relevant literature examining excess comovement.

<Insert Table 3 Here>

We then turn to the test of Hypothesis 2 (H2), which posits that institution-driven cross-stock information diffusion also amplifies excess comovement. To test this hypothesis, we estimate Equation (8), with the results for the Chinese and U.S. markets presented in Columns (2) and (5) of Table 3, respectively. As shown in the table, the coefficient on *Inst* is significantly positive in both markets (China: $\beta = 0.899, p < 0.01$; U.S.: $\beta = 0.243, p < 0.01$), thereby supporting Hypothesis 2. These results suggest that information diffusion driven by institutional investors plays a crucial role in driving excess comovement in both markets.

Finally, we examine Hypotheses 3a and 3b by estimating Equation (9). The regression results for the Chinese market are presented in Column (3). As shown, the coefficients for *Flow* and *Inst* are 3.577 and 0.776, respectively. Compared with the results in Columns (1) and (2), the coefficients slightly decrease but remain significantly positive at the 1% level. These findings indicate that both retail-driven information diffusion (*Flow*) and institution-driven information diffusion (*Inst*) have a significant impact on excess comovement, even when both variables are included in the same model. Additionally, the coefficient on *Flow* is nearly five times larger than that on *Inst*, and the coefficient difference test confirms that this difference is highly significant ($z = 56.32$). These findings provide strong evidence that, in the Chinese market, compared with institution-driven diffusion, retail-driven information diffusion has a substantially stronger effect on excess comovement, thus supporting Hypothesis 3a.

For the U.S. market, the regression results are presented in Column (6). The coefficients on *Flow* and *Inst* are 0.009 and 0.243, respectively. In comparison to the results in Columns (4) and (5), the coefficient on *Flow* decreases slightly, whereas the coefficient on *Inst* remains largely unchanged. Both coefficients remain significantly positive at the 1% level, suggesting that both retail- and institution-driven information diffusion contribute to excess comovement, even when analyzed together in the same model.

These findings are consistent with recent studies underscoring the growing influence of retail investors in the U.S. market, particularly on platforms such as StockTwits and Robinhood (Giannini et al., 2019; Cookson and Niessner, 2020; Cookson et al., 2024; Hirshleifer et al., 2025; Vamossy and Skog, 2025). Notably, the coefficient for *Inst* is nearly 27 times larger than that for *Flow*, and the coefficient difference test shows that this difference is highly significant ($z = 211.82$). These results suggest that in the U.S. market, institution-driven information diffusion has a far stronger effect on excess comovement than retail-driven diffusion does, thereby supporting Hypothesis 3b.

In summary, these results strongly support Hypotheses 1 through 3, indicating that investor-driven information diffusion has a significant and positive impact on excess comovement. Crucially, the relative economic importance of each information channel varies across markets with different structures: In China, retail investors play a dominant role in influencing excess comovement, whereas in the U.S., institutional investors are the primary drivers.

*4.2 Endogeneity Issues*

Recent studies show that common institutional investors strongly and robustly affect excess comovement (Anton and Polk, 2014; Koch et al., 2016; Geraci et al., 2023). Following these studies, we employ the propensity score matching (PSM) method and instrumental variable (IV) approach to address potential endogeneity concerns arising from the possibility that stocks with greater excess comovement are more likely to be held by common institutional investors. Our results align with these studies, showing that institution-driven information diffusion amplifies excess comovement in both the Chinese and U.S. markets. However, our finding that retail-driven information diffusion contributes to excess comovement may be subject to serious endogeneity concerns. Specifically, stocks exhibiting stronger comovement may attract more attention from retail investors, thereby potentially increasing the magnitude of retail-driven information diffusion among these stocks. To mitigate this concern, we implement multiple strategies including a quasi-natural experiment, propensity score matching (PSM), and instrumental variable (IV) analysis.

*4.2.1 Quasi-natural experiment*

To address endogeneity concerns and establish a causal relationship between information diffusion and excess comovement, we utilize quasi-natural experiments in both the Chinese and U.S. markets. Specifically, we focus on the exogenous shocks generated by the launches of the Eastmoney mobile application in China (November 2012) and the StockTwits mobile application in the U.S. (October 2010). Both events significantly increase investor interactions on their respective platforms—Guba in China and StockTwits in the U.S.—without directly affecting stock prices. These increases in forum activity provide an ideal setting to examine the causal impact of retail-driven information diffusion on excess comovement.

We compare the pre-event and post-event periods to observe changes in the impact of information diffusion on excess comovement. The results are presented in Table 4 (Panel A for the Chinese market and Panel B for the U.S. market). Specifically, Columns (1) and (2) highlight significant changes in the coefficient for *Flow* before and after the app launches. In the Chinese market, the coefficient for *Flow* increases from 11.933 in 2012Q3 (pre-event) to 28.737 in 2012Q4 (post-event), nearly doubling after the mobile app's release. Similarly, in the U.S. market, the coefficient for *Flow* increases from 0.031 in 2010Q3 (pre-event) to 0.064 in 2010Q4 (post-event), showing a substantial increase after the launch of the StockTwits mobile application. These results suggest that the launch of the mobile applications significantly amplifies the influence of retail-driven information diffusion on excess comovement in both markets.

<Insert Table 4 Here>

Moreover, to rule out the possibility that the observed effects are driven by seasonal trends or external temporal factors, we perform placebo tests by comparing the pre- and post-event periods across subsequent years in both markets. For China, we select the years 2015, 2018, and 2021 as placebo periods; for the U.S., the chosen placebo years are 2013, 2016, and 2019. The results of these placebo tests reveal no significant differences in the coefficients on *Flow* between the pre- and post-periods, thereby confirming that the observed effects can indeed be attributed to the app launches. Overall, these quasi-natural experiments provide robust evidence that increased retail investors' interactions on stock forums can significantly influence excess comovement, demonstrating a substantial impact of retail-driven information diffusion in both the Chinese and U.S. markets.

*4.2.2 Propensity score matching*

To further address potential endogeneity concerns and control for sample selection bias, we employ the propensity score matching (PSM) method. The PSM approach effectively mitigates the risk of bias arising from the non-random assignment of stock pairs to different levels of information diffusion, thereby ensuring a more balanced comparison between stocks with varying levels of retail- and institution-driven information diffusion.

We begin by categorizing stocks into different groups based on their *Flow* (retail-driven information diffusion) and *Inst* (institution-driven information diffusion) values. Specifically, we divide the sample into quartiles according to the levels of *Flow* and *Inst*, and then employ nearest-neighbor matching with a 1:1 matching ratio (without replacement) to match pairs from the highest quartile with those from other quartiles that exhibit similar propensity scores. To generate these propensity scores, we estimate a logistic regression model using the same control variables as in the baseline regressions. To increase the quality of matching, we impose a caliper of 0.01, ensuring that the difference in propensity scores between matched pairs does not exceed 1%. Although we do not report the mean differences in the matching variables due to space limitations, we confirm that there are no significant differences between the treatment and control groups across all the matching variables. This procedure ensures that the pairs being compared are more similar in terms of their observable characteristics, thereby reducing potential selection bias.

<Insert Table 5 Here>

After completing the matching procedure, we re-estimate Equation (9) using the matched sample. The results presented in Table 5 (Panel A for the Chinese market and Panel B for the U.S. market) show that, after the matching process, the coefficients for *Flow* and *Inst* remain significantly positive in both markets. These findings confirm that even after sample selection bias is addressed, both retail-driven and institution-driven information diffusion continue to significantly affect excess comovement.

*4.2.3 Instrumental variable approach*

To further mitigate potential endogeneity concerns, we employ the instrumental variable (IV) approach, which helps identify causal relationships by utilizing instruments that are correlated with the endogenous independent variables but uncorrelated with the error terms.

For retail-driven information diffusion in China, we utilize investor registration length (*Age*) as our instrumental variable. Investors with longer registration histories are more likely to have consistent and influential participation in stock forums. They tend to post and reply more frequently, thereby promoting information diffusion. Moreover, this variable is unlikely to directly affect excess comovement, satisfying the criteria for a valid instrument. As shown in Table 5 (Panel A, Column 3), the first-stage regression result reveals a strong positive relationship between investor age and retail-driven information diffusion, with an F-statistic of 149.255, indicating that the instrument is sufficiently strong and relevant. The second-stage regression further confirms that the instrumented *Flow* maintains a significant positive effect on excess comovement, with a coefficient of 2.561.

For retail-driven information diffusion in the U.S. market, we utilize the StockTwits app update event that occurred on September 6, 2013, as an instrument variable. This update introduced several new features, including push notifications, in-app notifications, and improved message views, which significantly strengthen real-time communication among retail investors on the platform. This update event is exogenous to the stock market and directly affects the intensity of retail-driven information diffusion on StockTwits, thereby making it an appropriate instrument for our analysis. We construct a dummy variable, *PostUpdate*, which equals 1 after the event and 0 before it, capturing the exogenous increase in retail-driven information diffusion resulting from the update. As shown in Table 5 (Panel B, Column 3), the coefficient for *PostUpdate* is 0.570, statistically significant at the 1% level, confirming that the app update significantly enhances information diffusion. In the second stage, we use the predicted *Flow* to examine its effect on excess comovement. The results presented in Column (4) of Panel B demonstrate that *Flow* has a significant impact on excess comovement in the U.S. market.

For institutional investors, we introduce merger events of financial institutions as an instrument to address the endogeneity problem in the context of institution-driven information diffusion in both the Chinese and U.S. markets. This instrument is chosen on the basis of its relevance and exogeneity. When a financial institution undergoes a merger, it typically consolidates its holdings across multiple firms, resulting in changes to common institutional ownership. Furthermore, the merger event is driven by strategic decisions of the merging institutions rather than by factors that directly affect the related stocks, thereby making it a suitable instrument. Accordingly, we utilize financial institution mergers (*Dmerger*) as our instrumental variable, defined as a dummy variable that takes a value of 1 during the quarter following the merger event and 0 otherwise. In the first stage, we estimate the impact of merger events (*Dmerger*) on common institutional ownership (*Inst*). As presented in Columns (5) of Panels A and B, the coefficients for *Dmerger* are statistically significant at the 1% level, indicating that merger events lead to increased institution-driven information diffusion. In the second stage, we use the predicted *Inst* to examine its effect on excess comovement. The results show that institution-driven information diffusion significantly impacts excess comovement in both the Chinese and U.S. markets, as expected.

To summarize, the results from the quasi-natural experiment, propensity score matching (PSM), and instrumental variable analysis alleviate concerns regarding endogeneity between information diffusion and excess comovement, confirming that both retail-driven and institution-driven information diffusion have a significant positive effect on excess comovement.

*4.3 Robustness Tests*

*4.3.1 Alternative comovement measures*

To validate the robustness of our findings and ensure that the observed effects are not sensitive to the measure of comovement, we conduct robustness tests using alternative proxies for excess comovement. Specifically, we substitute the Fama-French five-factor (FF5) model employed in the baseline analysis with two alternative models: the Fama-French three-factor (FF3) model and the capital asset pricing model (CAPM) (Sharpe, 1964).

<Insert Table 6 Here>

The results of the robustness tests are presented in Table 6 (Panel A for the Chinese market and Panel B for the U.S. market). In both panels, we report the regression results for comovement measures derived from FF3 residuals (Columns 1 and 5) and CAPM residuals (Columns 2 and 6). For both the Chinese and U.S. markets, we observe that the coefficients for *Flow* and *Inst* remain significantly positive, thereby confirming the robustness of our main results.

*4.3.2 Alternative institution-driven information diffusion measures*

To further ensure the robustness of our findings, we conduct additional tests utilizing alternative measures of institution-driven information diffusion. Specifically, we modify the threshold for institutional ownership to examine whether the relationship between institution-driven information diffusion and excess comovement remains consistent across different levels of institutional involvement.

The results of these robustness tests are presented in Table 6 (Panel A for the Chinese market and Panel B for the U.S. market). For institution-driven information diffusion, we test two alternative thresholds for common institutional ownership: 1% and 3%. These thresholds are defined as the proportion of shares held by institutions, where institutional ownership exceeds 1% or 3% of the total outstanding shares of a stock. These findings are consistent with our main results. Specifically, for the Chinese market, the coefficients for *Inst* (1%) and *Inst* (3%) remain significant at the 1% level. Similarly, in the U.S. market, the coefficient for institution-driven information diffusion remains significant at the 1% level across both thresholds, with values of 0.149 (1% threshold) and 0.111 (3% threshold).

*4.3.3 Excluding stocks from the same industry*

To account for potential industry-specific effects on excess comovement, we conduct an additional robustness check by excluding stock pairs from the same industry. The rationale for this test is based on the understanding that stocks within the same industry may exhibit greater comovement due to shared factors such as economic cycles, market trends, and industry-specific news. These factors could potentially confound the relationship between information diffusion and excess comovement. By excluding stocks within the same industry, we aim to isolate the effects of information diffusion from any industry-specific influences.

<Insert Table 7 Here>

The results of this analysis are presented in Column (1) of Table 7 (Panel A for the Chinese market and Panel B for the U.S. market). The coefficients for *Flow* and *Inst* remain significantly positive in both markets. In the Chinese market, the coefficient for *Flow* is 3.373, and for *Inst*, it is 0.348; both are statistically significant at the 1% level. Similarly, in the U.S. market, the coefficients for *Flow* and *Inst* are 0.009 and 0.233, respectively, which are also statistically significant at the 1% level. These results indicate that excluding stocks within the same industry does not substantially alter the observed effects of retail- and institution-driven information diffusion on excess comovement, thereby confirming that our findings are attributable to investor-driven information diffusion.

*4.3.4 High-dimensioned fixed effects*

To increase the precision of our estimates, we conduct an additional robustness check by incorporating high-dimensional fixed effects into the regression models. This approach accounts for interactions between stock and industry fixed effects, as well as time trends, to thereby capture both time-invariant and time-varying factors that may influence the relationship between information diffusion and excess comovement. High-dimensional fixed effects control for unobserved heterogeneity across stock pairs, industries, and time periods, thereby ensuring that our findings are not confounded by such factors.

The results of this analysis are presented in Column (2) of Table 7 (Panel A for the Chinese market and Panel B for the U.S. market). After incorporating high-dimensional fixed effects, we observe that the coefficients for *Flow* and *Inst* remain statistically significant at the 1% level. Specifically, in the Chinese market, the coefficient for *Flow* is 3.428 and for *Inst* is 0.231. In the U.S. market, the coefficients for *Flow* and *Inst* are 0.010 and 0.241, respectively, and both are significant at the 1% level. These findings confirm that the inclusion of high-dimensional fixed effects does not materially alter our results, indicating that the observed effects of information diffusion on excess comovement are robust even after we control for industry-specific and time-varying factors.

*4.3.5 Alternative measures of control variables*

To further strengthen the robustness of our findings, we conduct an additional robustness check by modifying the measurement of control variables. In the baseline analysis, we employ dummy variables to indicate whether the two stocks in a pair share similar characteristics, such as industry, size, and other financial metrics. In this section, we utilize continuous variables that capture the actual differences in these characteristics between stock pairs, moving beyond binary classifications. Specifically, we calculate the absolute value of the natural logarithm of the difference between stock $i$ and stock $j$ for each variable: price, size, return on assets, leverage ratio, volatility, and momentum. This approach enables a more granular examination of fundamental similarities between stocks, which might otherwise confound the relationship between information diffusion and excess comovement.

The results of this alternative measure are presented in Columns (3) of Table 7 (Panel A for the Chinese market and Panel B for the U.S. market). In this robustness check, we observe that the coefficients on *Flow* and *Inst* remain highly significant in both markets, consistent with our main findings. These results suggest that employing continuous variables as control measures does not materially alter the conclusions drawn from our baseline analysis, thereby demonstrating that the observed relationships are robust to alternative specifications of control variables.

*4.3.6 Excluding observations with low forum interactions*

To enhance the reliability of our findings and reduce potential noise in the measure of retail-driven information diffusion, we conduct a robustness check by excluding observations with low levels of forum interactions. Specifically, we remove stock pairs where the number of replies triggered by co-investors in the corresponding stock forums is fewer than three. By excluding these low-interaction pairs, we aim to focus on stocks that exhibit substantial engagement within stock forums, which are more likely to reflect genuine information diffusion rather than random or superficial interactions.

The results of this robustness check are reported in Column (4) of Table 7 (Panel A for the Chinese market and Panel B for the U.S. market). After excluding observations with low forum interaction, the coefficient for *Flow* remains significantly positive in both markets. These findings indicate that the exclusion of these observations does not materially affect the findings, thereby supporting the robustness of our conclusions.

*4.3.7 Fama-Macbeth regression*

To address potential issues related to cross-sectional dependence, we apply the Fama-Macbeth (FM) regression methodology. This approach is particularly suitable for our analysis as it accounts for the time-varying nature of the relationship between information diffusion and excess comovement. Furthermore, it generates unbiased standard errors that adjust for cross-sectional correlation, allowing us to obtain more reliable estimates over time. We begin by conducting cross-sectional regressions for each quarter, where the dependent variable is $Cor^{FF5}$, and the key independent variables are *Flow* and *Inst*, alongside a set of control variables. Each quarterly regression captures the relationship between information diffusion and excess comovement for all stock pairs within that specific quarter. Subsequently, we calculate the average of the estimated coefficients obtained from these quarterly regressions across the entire sample period. This averaging procedure yields final coefficient estimates that reflect the average effect of information diffusion on excess comovement across all quarters, while effectively addressing any cross-sectional dependence in the residuals.

The results of the FM regression are reported in Table 8, with Panel A corresponding to the Chinese market and Panel B to the U.S. market. In the Chinese market, the FM regression results indicate that *Flow* has a statistically significant and positive impact on excess comovement, thereby corroborating the conclusions derived from the baseline regressions. The coefficient for *Inst* is also statistically significant at the 1% level, further supporting the relationship between institution-driven information diffusion and excess comovement in China.

<Insert Table 8 Here>

Similarly, in the U.S. market, the results of the FM regression reveal that both *Flow* and *Inst* exert significant and positive effects on excess comovement, which is consistent with our main findings. In addition to *Flow* and *Inst*, these results also confirm the significance of other control variables, including news and analyst coverage, and key fundamental characteristics. These variables exhibit consistent and significant effects on excess comovement, as reflected in their coefficients, which align with prior research and further substantiate the robustness of our findings.

*4.4 Mechanism analysis*

To empirically examine the mechanism through which information diffusion influences excess comovement, we adopt the methodology employed in recent studies (Jiang et al., 2022; Gong et al., 2024; Safiullah et al., 2024; Chen et al., 2024) to investigate the mediating role of investors' trading behavior in the relationship between information diffusion and excess comovement. This analysis tests Hypothesis 4, which proposes that information diffusion affects excess comovement by promoting investors' trading behavior.

The results of the mechanism analysis are reported in Table 9, with Panel A corresponding to the Chinese market and Panel B to the U.S. market. In both panels, Columns (2) and (4) demonstrate that the coefficients for *Flow* and *Inst* are statistically significant at the 1% level, indicating a strong positive association between information diffusion and investors' trading behavior in both markets. These results support the notion that information diffusion driven by both retail and institutional investors significantly promotes their trading decisions.

<Insert Table 9 Here>

To examine whether investors' trading behavior can explain the impact of information diffusion on excess comovement, we extend the regression models by incorporating the buy-sell imbalance (BSI) as a proxy for trading behavior. The results are reported in Columns (3) and (5) of Panel A and Panel B, respectively. In both the Chinese and U.S. markets, when the correlation of the buy-sell imbalance for retail (institutional) investors, $COR_{BSI}^{Retl}$ ($COR_{BSI}^{Inst}$), is included in the regression, the coefficient for *Flow* (*Inst*) decreases, indicating that retail (institutional) investors' trading behavior plays a significant mediating role in the relationship between information diffusion and excess comovement. Overall, these results provide support for Hypothesis 4, which proposes that information diffusion affects excess comovement through investors' trading behavior.

*4.5 Predicting excess comovement*

*4.5.1 Lead-lag relationship of excess comovement*

Our baseline results indicate that retail-driven information diffusion significantly amplifies excess comovement. To further explore this phenomenon, we examine whether a lead-lag relationship exists in excess comovement between stocks characterized by fast versus slow retail-driven information diffusion in the Chinese and U.S. markets. The literature on stock return predictability investigates lead-lag patterns in stock returns, attributing such dynamics to informed trading. Research suggests that stocks with faster information diffusion attract more informed investors, thereby accelerating price reactions to new information (Holden and Subrahmanyam,1992; Pantzalis and Wang, 2017). Brennan et al. (1993) find that the returns of stocks followed by numerous analysts lead those followed by fewer analysts. Pantzalis and Wang (2017) define shareholder coordination as the weighted average of the geographical distance among a firm's institutional shareholders, proposing that it functions as an effective channel for information diffusion. Their findings show that stock returns for firms with high shareholder coordination precede those for firms with low shareholder coordination.

Based on these studies, we conjecture that stocks with rapid information diffusion respond to new information more quickly than those with slow information diffusion do. As a result, excess comovement among fast-diffusion stocks should precede that among slow-diffusion stocks. To test this conjecture, we adopt the methodology proposed by Pantzalis and Wang (2017), which involves calculating the information diffusion-weighted average correlation coefficient and comparing these coefficients between groups of fast- and slow-diffusion stocks. The calculation and analysis are carried out as follows. Using monthly data from 2010 to 2022, we first compute the excess comovement for all stock pairs by measuring the within-month correlation of daily excess returns. Based on the level of retail investors' monthly information diffusion, stock pairs are classified into five groups, ranging from fastest to slowest diffusion. For each group, we calculate the weighted average correlation for each quintile, where the weights represent each pair's proportion of information diffusion relative to the total within its group.

Table 10 presents the mean and median of each group's weighted average correlation coefficient over the period 2010–2022. Panel A indicates that, in the Chinese stock market, the mean (median) correlation of the fast group significantly exceeds that of slow group by 0.0136 (0.0166) ($p < 0.01$), thereby confirming a stronger contemporaneous excess comovement among fast-diffusion stocks. Panel B reveals that, in the U.S. stock market, while there is a positive difference between the fast and slow groups, it is significant only at the 10% level, suggesting a weaker differentiation between these groups.

<Insert Table 10 Here>

Given that the excess comovement of the group with fast information diffusion is stronger than that of the group with slow information diffusion in the current period, we propose that this difference arises because stocks in the fast information diffusion group respond to new information more rapidly. Accordingly, we further examine the lead-lag relationship of excess comovement through linear regression analysis. As faster information diffusion facilitates quicker price reactions to new information, the excess comovement of stock pairs with fast information diffusion is expected to lead that of pairs with slow information diffusion. We test this hypothesis using the following regression model:

$$COR_t^{slow} = c_0 + c_1 COR_{t-1}^{fast} + \varepsilon, \quad (10)$$

where $COR_{t-1}^{fast}$ ($COR_t^{slow}$) denotes the information diffusion-weighted average correlation coefficient for the group with fast (slow) information diffusion in month $t-1$ ($t$).

<Insert Table 11 Here>

The results of estimating Equation (10) are presented in Table 11. In the Chinese stock market, the average correlation coefficient of the group with fast information diffusion ($COR_{t-1}^{fast}$) in month $t-1$ significantly leads that of the group with slow information diffusion in month $t$, with an estimated value of 0.2913 that is statistically significant at the 1% level. In the U.S. stock market, $COR_{t-1}^{fast}$ is 0.0821, which is statistically significant at the 10% level, indicating a weaker and marginally significant effect. Collectively, these findings highlight a strong lead-lag effect in China, in contrast to the more muted results observed in the U.S., which is consistent with our core conclusion that retail-driven information diffusion predominantly influences excess comovement in retail-oriented markets.

*4.5.2 Prediction effect of information diffusion on excess comovement*

In the previous section, we establish a lead-lag relationship between information diffusion and excess comovement, demonstrating that excess comovement within the group of stocks with fast information diffusion precedes that of stocks with slow information diffusion. These findings indicate that stocks with faster information diffusion respond to new information more rapidly than those with slower information diffusion do. Building on these insights, we now extend our analysis by examining the predictability of both retail- and institution-driven information diffusion on excess comovement. We investigate the effects of lagged information diffusion over up to four periods. This extended analysis enables us to assess the long-term impact of both retail and institutional information diffusion on excess comovement. The regression model used for this analysis is as follows:

$$Cor_{ij,t} = \rho_0 + \sum_{n=1}^{4} \rho_n Flow_{ij,t-n} + \sum_{n=1}^{4} \rho_n Inst_{ij,t-n} + \sum_{n=1}^{4} \rho_n Control_{ij,t-n} + \epsilon, \quad (11)$$

where $Cor_{ij,t}$ is the within-quarter realized correlation of daily excess returns for stock pair $i$ and $j$ in quarter $t$; $Flow_{ij,t-n}$and $Inst_{ij,t-n}$ represent the lagged retail- and institution-driven information diffusion for stock pair $i$ and $j$ in quarter $t-n$, respectively; and $Control_{ij,t-n}$ represents the control variables for stock pair $i$ and $j$ in quarter $t-n$, identical to the control variables in subsection 4.1 but adjusted for each lag period. By including lagged variables for *Flow*, *Inst*, and control variables across periods ranging from 1 to 4 lags, this model effectively captures the impact of information diffusion and other control variables over time on excess comovement. Table 12 presents the regression results of Equation (11) for both markets.

The results for the Chinese market (Panel A, Columns 1–4) reveal distinct predictive patterns between the two investor types. Retail-driven information diffusion strongly persists. The coefficient on *L. Flow* (3.459, $p < 0.01$) confirms significant predictability one quarter ahead. Notably, this predictive power remains robust across longer lags: *L2. Flow* (1.419, $p < 0.01$), *L3. Flow* (1.420, $p < 0.01$), and *L4. Flow* (1.321, $p < 0.01$) are all highly significant. The magnitude of *L4. Flow* underscores the lasting impact of retail-driven information diffusion in this market. In contrast, institution-driven diffusion shows limited persistence. While *L. Inst* is statistically significant (0.259, $p < 0.01$), its effect size is considerably smaller than that of *L. Flow*. More importantly, the predictive power declines rapidly over time: *L2. Inst* shows marginal significance only in certain specifications (0.134, $p < 0.05$ in Column 2), and *L3. Inst* and *L4. Inst* are statistically insignificant. This sharp divergence highlights that only retail-driven diffusion has significant multi-quarter predictive power in the Chinese retail-dominated market.

The results for the U.S. market (Panel B, Columns 5–8) reveal an inverse relationship consistent with the prevailing institutional dominance. Institution-driven diffusion exhibits remarkably persistent predictability. The coefficients are highly significant across all lags: *L. Inst* (0.191, $p < 0.01$), *L2. Inst* (0.109, $p < 0.01$), *L3. Inst* (0.073, $p < 0.01$), and *L4. Inst* (0.027, $p < 0.01$). The sustained significance of *L4. Inst* underscores the long-lasting influence of institutional investors. In contrast, retail-driven diffusion exhibits only transient predictive power. *L. Flow* is significant (0.012, $p < 0.01$), but this effect diminishes rapidly, as evidenced by the statistical insignificance of *L2. Flow*, *L3. Flow*, and *L4. Flow* in subsequent periods. These findings confirm that institutional diffusion underpins multi-quarter predictability in the U.S., whereas retail diffusion provides only short-term signals.

<Insert Table 12 Here>

These lagged effects strongly reinforce the dominance of the primary investor type in shaping predictability. In China, retail-driven diffusion (*Flow*) exerts a strong and persistent predictive influence on excess comovement across multiple quarters, far surpassing the weak and transient predictive role of institution-driven diffusion (*Inst*). In contrast, in the U.S., institution-driven diffusion (*Inst*) provides robust multi-quarter predictability, whereas retail-driven diffusion (*Flow*) offers only fleeting predictive power, limited to a single quarter. This marked difference in the persistence of predictive power—where retail-driven effects endure in China and institution-driven effects prevail in the U.S. —directly reflects the underlying investor structures and their associated information diffusion channels documented in our baseline findings (Section 4.1).

## 5. Conclusion

This study examines how cross-stock information diffusion, driven by both retail and institutional investors, influences excess comovement in two structurally distinct markets: the retail-dominated Chinese market and the institution-dominated U.S. market. Our sample covers major exchanges in both markets (NYSE, NASDAQ, and AMEX for the U.S.; SSE and SZSE for China) from 2010 to 2022. Through rigorous identification strategies and comprehensive robustness checks, our findings yield the following key insights: First and most critically, the dominant investor group in each market exerts the most substantial influence on excess comovement. In the Chinese retail-dominated market, compared with institution-driven diffusion, retail-driven information diffusion has a significantly stronger effect on excess comovement. In contrast, in the U.S. market, institution-driven information diffusion plays a more prominent role in influencing excess comovement, surpassing the impact of retail-driven diffusion. Second, we identify investors' trading behavior as the mechanism through which information diffusion affects excess comovement. In both markets, retail- and institution-driven information diffusion significantly amplifies investors' trading behavior. This synchronized trading behavior, in turn, mediates the influence of information diffusion on excess comovement, thereby supporting the theoretical premise that information is incorporated into prices primarily through investor transactions. Third, the predictive analysis reveals the presence of a lead-lag relationship and persistent investor-specific effects. We find that stocks with faster retail-driven information diffusion exhibit excess comovement that precedes those with slower diffusion, confirming the causal direction of this relationship. Importantly, the persistence of this predictive power aligns closely with the structural characteristics of each market: In China, retail-driven diffusion demonstrates a strong, multi-quarter predictive effect on excess comovement, significantly outperforming the weak and transient influence of institutional diffusion. In contrast, in the U.S. market, institutional diffusion exhibits robust predictability over multiple quarters, whereas retail diffusion offers only limited and transient predictive power, confined primarily to a single quarter.

Our study makes several contributions to the existing literature. First, we introduce an information consumption framework that complements prevailing supply- and demand-side explanations for excess comovement. By tracing retail investors' information interactions across subforums, we reveal the pathways of information diffusion among related stocks, providing a more direct and accurate measure of cross-stock information channels than previous proxies. Second, we conduct a joint analysis of both retail and institutional information diffusion channels, moving beyond studies that focus exclusively on a single investor type. This comprehensive analysis reveals the distinct effects of retail- and institution-driven information diffusion on excess comovement. Third, we present the first empirical study that simultaneously examines investor-driven information diffusion in two structurally distinct markets: the retail-dominated Chinese market and the institution-dominated U.S. market. Finally, we provide empirical evidence on investors' trading behavior as a key mechanism through which information diffusion influences excess comovement, thereby responding to a key call in the literature.

Our findings offer practical implications for both market participants and regulators in the Chinese and U.S. markets. In the U.S. market, institutional investors should systematically monitor information embedded within ownership networks. By tracking shared events (e.g., supply-chain disruptions, regulatory changes) among firms within their portfolios, they can proactively adjust their investment positions to mitigate synchronized downside risks arising from excess comovement. Regulators may consider increasing the transparency of institutional ownership by requiring more frequent disclosures of stake changes, thereby enabling market participants to better assess comovement-related risks. In the Chinese stock market, given the predictable impact of retail-driven information diffusion on excess comovement, retail investors may benefit from developing trading strategies that incorporate insights derived from discussions on stock forums related to specific stocks to enhance portfolio returns. Regulators should closely monitor activities on online stock forums and implement preventive measures to mitigate risks—such as monitoring and countering the spread of false or misleading information—and encourage firms to promptly address market rumors or disclose relevant information in a timely manner.

Our study has two limitations that suggest potential directions for future research. First, our measure of retail-driven information diffusion is based on the volume of posts and replies. This metric could be enhanced through the application of textual analysis techniques to extract meaningful insights from the content of these interactions. Future research could adopt a qualitative textual similarity approach to identify thematic commonalities in information interactions, thereby enabling a more refined and insightful assessment of how retail-driven information diffusion influences excess comovement. Second, although Guba and StockTwits are widely recognized as representative platforms for dedicated stock-related discussions within their respective markets, the growing influence of broader social media platforms (e.g., Reddit, X) on the stock market cannot be overlooked. Future research could extend our framework by incorporating multi-platform data sources and conducting systematic comparisons of information diffusion patterns across specialized investment forums and general-purpose social media platforms. This would provide a more comprehensive understanding of how diverse digital environments shape investor behavior and market dynamics.

**Table 1**
**Descriptive statistics of retail investor activity on stock forums.**

| | Chinese stock market | | | | U.S. stock market | | | |
|---|---|---|---|---|---|---|---|---|
| **Year** | **No. Stocks** | **No. Stockpairs** | **No. Co-investors Per Pair** | **No. Posts and Replies Per Pair** | **No. Stocks** | **No. Stockpairs** | **No. Co-investors Per Pair** | **No. Posts and Replies Per Pair** |
| 2010 | 1,416 | 986,820 | 0.90 | 1.94 | 1,108 | 579,726 | 3.50 | 33.36 |
| 2011 | 1,753 | 1,505,628 | 1.39 | 3.44 | 1,160 | 670,717 | 8.01 | 101.07 |
| 2012 | 2,011 | 1,992,491 | 1.41 | 2.86 | 1,772 | 1,561,715 | 9.08 | 86.89 |
| 2013 | 2,193 | 2,385,063 | 2.01 | 6.62 | 2,142 | 2,275,435 | 8.61 | 63.30 |
| 2014 | 2,257 | 2,408,525 | 3.43 | 20.17 | 2,317 | 2,670,405 | 15.09 | 97.61 |
| 2015 | 2,351 | 2,709,831 | 7.38 | 60.91 | 2,586 | 3,334,107 | 15.44 | 202.93 |
| 2016 | 2,541 | 3,203,932 | 20.10 | 138.18 | 2,729 | 3,715,283 | 17.41 | 216.10 |
| 2017 | 2,827 | 3,986,762 | 25.27 | 165.10 | 2,925 | 4,260,731 | 38.95 | 417.42 |
| 2018 | 3,203 | 5,125,420 | 22.21 | 627.92 | 3,120 | 4,855,662 | 42.34 | 528.37 |
| 2019 | 3,275 | 5,360,544 | 29.28 | 719.55 | 3,303 | 5,442,845 | 40.90 | 493.07 |
| 2020 | 3,487 | 6,021,382 | 39.73 | 586.77 | 3,669 | 6,724,470 | 52.28 | 605.17 |
| 2021 | 4,012 | 7,949,559 | 47.15 | 497.92 | 4,210 | 8,744,346 | 70.41 | 647.33 |
| 2022 | 4,533 | 10,258,480 | 34.80 | 320.26 | 4,517 | 10,141,971 | 42.54 | 631.91 |

This table presents key metrics reflecting dynamics of retail investors' interactions on China's Eastmoney (covering 4,533 stocks) and U.S. StockTwits (covering 4,517 stocks) platforms from 2010 to 2022. The statistics reported include: (i) the number of stocks actively discussed, (ii) the number of stock pairs, (iii) the number of co-investors per stock pair, and (iv) the number of posts and replies per stock pair.

**Table 2**
**Descriptive statistics of variables.**

**Panel A: Chinese stock market**

| Variable | N | Mean | SD | Min | P25 | Median | P75 | Max |
|---|---|---|---|---|---|---|---|---|
| $Cor^{FF5}$ | 179,506,286 | 0.007 | 0.154 | –1 | –0.096 | 0.003 | 0.106 | 1 |
| *Flow* | 179,506,286 | 106.935 | 281.520 | 0 | 16 | 54 | 133 | 194,961 |
| *Inst* | 179,506,286 | 0.663 | 6.485 | 0 | 0 | 0 | 0.006 | 60.656 |
| *Z_Flow* | 179,506,286 | 0 | 1 | –0.380 | –0.323 | –0.188 | 0.093 | 692.149 |
| *Z_Inst* | 179,506,286 | 0 | 1 | –0.102 | –0.102 | –0.102 | –0.101 | 9.353 |
| $COR_{BSI}^{Inst}$ | 179,506,286 | 0.073 | 0.157 | –1 | –0.032 | 0.070 | 0.174 | 1 |
| $COR_{BSI}^{Retl}$ | 179,506,286 | 0.082 | 0.165 | –1 | –0.029 | 0.076 | 0.186 | 1 |
| *News* | 179,506,286 | 0.031 | 1.398 | 0 | 0 | 0 | 0 | 606 |
| *Analyst* | 179,506,286 | 0.009 | 0.160 | 0 | 0 | 0 | 0 | 34 |
| *Similarroa* | 179,506,286 | 0.100 | 0.300 | 0 | 0 | 0 | 0 | 1 |
| *Similarlever* | 179,506,286 | 0.100 | 0.300 | 0 | 0 | 0 | 0 | 1 |
| *Similarsize* | 179,506,286 | 0.100 | 0.301 | 0 | 0 | 0 | 0 | 1 |
| *Similarprice* | 179,506,286 | 0.1 | 0.300 | 0 | 0 | 0 | 0 | 1 |
| *Similarmom* | 179,506,286 | 0.100 | 0.300 | 0 | 0 | 0 | 0 | 1 |
| *Similarvol* | 179,506,286 | 0.100 | 0.301 | 0 | 0 | 0 | 0 | 1 |
| *Sameind* | 179,506,286 | 0.444 | 0.497 | 0 | 0 | 0 | 1 | 1 |
| *Samelocation* | 179,506,286 | 0.078 | 0.268 | 0 | 0 | 0 | 0 | 1 |

**Panel B: U.S. stock market**

| Variable | N | Mean | SD | Min | P25 | Median | P75 | Max |
|---|---|---|---|---|---|---|---|---|
| $Cor^{FF5}$ | 139,293,862 | 0.078 | 0.393 | –1 | –0.113 | 0.203 | 0.493 | 1 |
| *Flow* | 139,293,862 | 172.619 | 408.702 | 0 | 50 | 153 | 486 | 622,955 |
| *Inst* | 139,293,862 | 48.682 | 45.897 | 0 | 49.253 | 38.159 | 83.387 | 143.919 |
| *Z_Flow* | 139,293,862 | 0 | 1 | –0.130 | –0.096 | –0.066 | –0.020 | 844.326 |
| *Z_Inst* | 139,293,862 | 0 | 1 | –1.061 | –0.953 | –0.229 | 0.756 | 30.296 |
| $COR_{BSI}^{Inst}$ | 139,293,862 | 0.004 | 0.197 | –1 | –0.102 | 0.004 | 0.112 | 1 |
| $COR_{BSI}^{Retl}$ | 139,293,862 | 0.003 | 0.133 | –1 | –0.087 | 0.001 | 0.092 | 1 |
| *News* | 139,293,862 | 0.018 | 0.103 | 0 | 0 | 0 | 0 | 104 |
| *Analyst* | 139,293,862 | 0.043 | 0.446 | 0 | 0 | 0 | 0 | 41 |
| *Similarroa* | 139,293,862 | 0.100 | 0.300 | 0 | 0 | 0 | 0 | 1 |
| *Similarlever* | 139,293,862 | 0.105 | 0.306 | 0 | 0 | 0 | 0 | 1 |
| *Similarsize* | 139,293,862 | 0.101 | 0.301 | 0 | 0 | 0 | 0 | 1 |
| *Similarprice* | 139,293,862 | 0.101 | 0.302 | 0 | 0 | 0 | 0 | 1 |
| *Similarmom* | 139,293,862 | 0.101 | 0.302 | 0 | 0 | 0 | 0 | 1 |
| *Similarvol* | 139,293,862 | 0.101 | 0.302 | 0 | 0 | 0 | 0 | 1 |
| *Sameind* | 139,293,862 | 0.169 | 0.255 | 0 | 0 | 0 | 0 | 1 |
| *Samelocation* | 139,293,862 | 0.333 | 0.471 | 0 | 0 | 0 | 1 | 1 |

This table presents the descriptive statistics for key variables in the Chinese (Panel A) and U.S. (Panel B) stock markets. The reported statistics include the mean, median, standard deviation, minimum, maximum, 25th percentile (P25), and 75th percentile (P75) values for each variable.

**Table 3**
**Baseline results.**

| | **Panel A: Chinese stock market** | | | **Panel B: U.S. stock market** | | |
|---|---|---|---|---|---|---|
| | (1) | (2) | (3) | (4) | (5) | (6) |
| | $Cor^{FF5}$ | $Cor^{FF5}$ | $Cor^{FF5}$ | $Cor^{FF5}$ | $Cor^{FF5}$ | $Cor^{FF5}$ |
| *Flow* | 3.598*** | | 3.577*** | 0.010*** | | 0.009*** |
| | (82.00) | | (81.90) | (23.90) | | (23.23) |
| *Inst* | | 0.899*** | 0.776*** | | 0.243*** | 0.243*** |
| | | (38.14) | (36.46) | | (240.02) | (239.88) |
| *News* | 0.002*** | 0.002*** | 0.002*** | 0.169*** | 0.065*** | 0.065*** |
| | (13.76) | (13.87) | (13.76) | (331.13) | (98.19) | (98.15) |
| *Analyst* | 0.032*** | 0.033*** | 0.032*** | 0.023*** | 0.023*** | 0.023*** |
| | (143.53) | (141.41) | (143.47) | (291.96) | (287.44) | (286.91) |
| *Similarroa* | 0.003*** | 0.003*** | 0.003*** | 0.018*** | 0.017*** | 0.017*** |
| | (70.93) | (73.85) | (70.56) | (189.10) | (181.05) | (180.91) |
| *Similarlever* | 0.003*** | 0.003*** | 0.003*** | 0.007*** | 0.007*** | 0.007*** |
| | (82.60) | (83.85) | (82.45) | (70.60) | (67.20) | (67.18) |
| *Similarsize* | 0.004*** | 0.004*** | 0.004*** | 0.021*** | 0.019*** | 0.019*** |
| | (96.58) | (112.05) | (93.65) | (208.43) | (188.58) | (188.29) |
| *Similarprice* | 0.008*** | 0.009*** | 0.008*** | 0.026*** | 0.024*** | 0.024*** |
| | (215.15) | (221.52) | (214.66) | (256.89) | (236.63) | (236.52) |
| *Similarmom* | 0.012*** | 0.012*** | 0.012*** | 0.172*** | 0.171*** | 0.171*** |
| | (296.80) | (304.42) | (296.55) | (1,702.09) | (1,699.22) | (1,699.06) |
| *Similarvol* | 0.003*** | 0.004*** | 0.003*** | 0.015*** | 0.015*** | 0.015*** |
| | (80.01) | (98.49) | (79.22) | (152.84) | (149.72) | (149.54) |
| *Samelocation* | 0.005*** | 0.005*** | 0.005*** | 0.005*** | 0.004*** | 0.004*** |
| | (104.18) | (105.54) | (103.75) | (44.06) | (36.06) | (36.02) |
| *Sameind* | 0.024*** | 0.024*** | 0.024*** | 0.051*** | 0.051*** | 0.051*** |
| | (514.34) | (513.78) | (514.29) | (412.10) | (408.06) | (408.04) |
| *Constant* | –0.007*** | –0.008*** | –0.007*** | 0.146*** | 0.147*** | 0.147*** |
| | (–292.20) | (–303.83) | (–291.25) | (2,712.20) | (2,722.05) | (2,722.14) |
| *Wald Test* | | | 2.801*** | | | 0.234*** |
| | | | (56.32) | | | (211.82) |
| Stock *i*× Quarter FE | Yes | Yes | Yes | Yes | Yes | Yes |
| Stock *j*× Quarter FE | Yes | Yes | Yes | Yes | Yes | Yes |
| Observations | 179,505,839 | 179,505,839 | 179,505,839 | 139,293,862 | 139,293,862 | 139,293,862 |
| Adjusted R2 | 0.033 | 0.033 | 0.033 | 0.393 | 0.393 | 0.393 |

This table reports the regression results examining the impact of investor-driven information diffusion on excess comovement in the two markets. The dependent variable, $Cor^{FF5}$, represents the correlation of daily Fama-French five-factor residual returns for each stock pair during the given quarter. The independent variables, *Flow* and *Inst*, are standardized to eliminate dimensional differences between them and ensure the comparability of the results. Definitions of other variables are reported in Appendix A. The individual and time fixed effects are included in the regression model. Standard errors are clustered at the stock-pair level. The t-statistics are reported in parentheses. ***, **, and * denote statistical significance at the 1%, 5%, and 10% levels, respectively.

**Table 4**
**Results of the quasi-natural experiment.**

**Panel A: Chinese stock market**

| | 2012Q3 | 2012Q4 | 2015Q3 | 2015Q4 | 2018Q3 | 2018Q4 | 2021Q3 | 2021Q4 |
|---|---|---|---|---|---|---|---|---|
| | (1) | (2) | (3) | (4) | (5) | (6) | (7) | (8) |
| | $Cor^{FF5}$ | $Cor^{FF5}$ | $Cor^{FF5}$ | $Cor^{FF5}$ | $Cor^{FF5}$ | $Cor^{FF5}$ | $Cor^{FF5}$ | $Cor^{FF5}$ |
| *Flow* | 11.933*** | 28.737*** | 9.483*** | 9.362*** | 4.792*** | 4.773*** | 2.800*** | 2.985*** |
| | (3.76) | (7.31) | (17.05) | (22.02) | (28.02) | (29.08) | (33.42) | (34.85) |
| *Inst* | 0.478*** | 0.522*** | 5.321*** | 5.659*** | 1.203*** | 2.270*** | 0.119*** | 0.682*** |
| | (3.03) | (3.89) | (16.26) | (19.01) | (6.72) | (12.59) | (3.98) | (12.14) |
| *Controls* | Yes | Yes | Yes | Yes | Yes | Yes | Yes | Yes |
| Stock *i* ×Quarter FE | Yes | Yes | Yes | Yes | Yes | Yes | Yes | Yes |
| Stock *j* ×Quarter FE | Yes | Yes | Yes | Yes | Yes | Yes | Yes | Yes |
| Observations | 532,857 | 534,839 | 2,598,088 | 2,943,322 | 5,195,411 | 5,192,188 | 9,890,052 | 10,113,940 |
| Adjusted R2 | 0.049 | 0.056 | 0.017 | 0.018 | 0.016 | 0.027 | 0.036 | 0.021 |

**Panel B: U.S. stock market**

| | 2010Q3 | 2010Q4 | 2013Q3 | 2013Q4 | 2016Q3 | 2016Q4 | 2019Q3 | 2019Q4 |
|---|---|---|---|---|---|---|---|---|
| | (1) | (2) | (3) | (4) | (5) | (6) | (7) | (8) |
| | $Cor^{FF5}$ | $Cor^{FF5}$ | $Cor^{FF5}$ | $Cor^{FF5}$ | $Cor^{FF5}$ | $Cor^{FF5}$ | $Cor^{FF5}$ | $Cor^{FF5}$ |
| *Flow* | 0.031*** | 0.064*** | 0.081*** | 0.046*** | 0.010*** | 0.003*** | 0.022*** | 0.015*** |
| | (4.09) | (8.25) | (7.94) | (4.63) | (8.12) | (3.21) | (9.98) | (7.21) |
| *Inst* | 0.258*** | 0.209*** | 0.169*** | 0.089*** | 0.036*** | 0.013*** | 0.185*** | 0.131*** |
| | (33.67) | (31.68) | (56.65) | (27.52) | (11.21) | (3.97) | (58.26) | (46.64) |
| *Controls* | Yes | Yes | Yes | Yes | Yes | Yes | Yes | Yes |
| Stock *i* ×Quarter FE | Yes | Yes | Yes | Yes | Yes | Yes | Yes | Yes |
| Stock *j* ×Quarter FE | Yes | Yes | Yes | Yes | Yes | Yes | Yes | Yes |
| Observations | 348,547 | 439,757 | 1,615,144 | 1,693,783 | 2,795,954 | 2,951,146 | 4,660,118 | 5,130,514 |
| Adjusted R2 | 0.412 | 0.216 | 0.478 | 0.139 | 0.096 | 0.093 | 0.144 | 0.088 |

This table presents regression results from the quasi-natural experiment of mobile app launches. The definitions of the variables are consistent with those in Equation (9). The individual and time fixed effects are included in the regression model. Standard errors are clustered at the stock-pair level. The t-statistics are reported in parentheses. ***, **, and * denote statistical significance at the 1%, 5%, and 10% levels, respectively.

**Table 5**
**Results of propensity score matching and instrumental variable analysis.**

**Panel A: Chinese stock market**

| | PSM | | IV-2SLS | | | |
|---|---|---|---|---|---|---|
| | **Retail** | **Institutional** | **Retail** | | **Institutional** | |
| | (1) | (2) | (3) | (4) | (5) | (6) |
| | $Cor^{FF5}$ | $Cor^{FF5}$ | *Flow* | $Cor^{FF5}$ | *Inst* | $Cor^{FF5}$ |
| *Flow* | 2.091*** | 1.916*** | | 2.561*** | | 2.454*** |
| | (32.52) | (11.80) | | (34.21) | | (24.71) |
| *Inst* | 0.406*** | 0.814*** | | 0.622*** | | 0.481*** |
| | (3.97) | (7.51) | | (16.28) | | (38.88) |
| *IV* | | | 0.780*** | | 0.314*** | |
| | | | (62.92) | | (51.29) | |
| *Controls* | Yes | Yes | Yes | Yes | Yes | Yes |
| Stock *i* ×Quarter FE | Yes | Yes | Yes | Yes | Yes | Yes |
| Stock *j* ×Quarter FE | Yes | Yes | Yes | Yes | Yes | Yes |
| Observations | 70,325,361 | 70,288,662 | 179,506,286 | 179,506,286 | 174,560,179 | 174,560,179 |
| *F-Statistic* | | | 149.255*** | | 215.853*** | |
| Adjusted R2 | 0.369 | 0.444 | 0.031 | 0.002 | 0.004 | 0.004 |

**Panel B: U.S. stock market**

| | PSM | | IV-2SLS | | | |
|---|---|---|---|---|---|---|
| | **Retail** | **Institutional** | **Retail** | | **Institutional** | |
| | (1) | (2) | (3) | (4) | (5) | (6) |
| | $Cor^{FF5}$ | $Cor^{FF5}$ | *Flow* | $Cor^{FF5}$ | *Inst* | $Cor^{FF5}$ |
| *Flow* | 0.004*** | 0.003 | | 0.093*** | | 0.107*** |
| | (8.68) | (0.54) | | (56.21) | | (20.89) |
| *Inst* | 0.153*** | 0.293*** | | 0.373*** | | 4.398*** |
| | (25.64) | (5.83) | | (212.94) | | (36.54) |
| *IV* | | | 0.570*** | | 0.213*** | |
| | | | (134.48) | | (42.23) | |
| *Controls* | Yes | Yes | Yes | Yes | Yes | Yes |
| Stock *i* ×Quarter FE | Yes | Yes | Yes | Yes | Yes | Yes |
| Stock *j* ×Quarter FE | Yes | Yes | Yes | Yes | Yes | Yes |
| Observations | 51,254,991 | 51,068,657 | 139,293,862 | 139,293,862 | 139,293,862 | 139,293,862 |
| *F-Statistic* | | | 272.192*** | | 385.264*** | |
| Adjusted R2 | 0.471 | 0.426 | 0.013 | 0.013 | 0.622 | 0.592 |

This table presents results from complementary strategies to address endogeneity: propensity score matching (PSM) and instrumental variables (IV). The definitions of the variables are consistent with those in Equation (9). The individual and time fixed effects are included in the regression model. Standard errors are clustered at the stock-pair level. The t-statistics are reported in parentheses. ***, **, and * denote statistical significance at the 1%, 5%, and 10% levels, respectively.

**Table 6**
**Robustness to alternative measures of excess comovement and institutional ownership.**

| | Panel A: Chinese stock market | | | | Panel B: U.S. stock market | | | |
|---|---|---|---|---|---|---|---|---|
| | **FF3 residual** | **CAPM residual** | **1% holding** | **3% holding** | **FF3 residual** | **CAPM residual** | **1% holding** | **3% holding** |
| | (1) | (2) | (3) | (4) | (5) | (6) | (7) | (8) |
| | $Cor^{FF3}$ | $Cor$ | $Cor^{FF5}$ | $Cor^{FF5}$ | $Cor^{FF3}$ | $Cor$ | $Cor^{FF5}$ | $Cor^{FF5}$ |
| *Flow* | 3.835*** | 5.649*** | 3.588*** | 3.594*** | 0.010*** | 0.014*** | 0.009*** | 0.010*** |
| | (81.77) | (87.11) | (81.95) | (81.98) | (23.48) | (37.48) | (23.16) | (23.33) |
| *Inst* (1%) | | | 0.430*** | | | | 0.149*** | |
| | | | (27.15) | | | | (166.31) | |
| *Inst* (3%) | | | | 0.264*** | | | | 0.111*** |
| | | | | (18.47) | | | | (121.50) |
| *Inst* | 0.990*** | 2.400*** | | | 0.242*** | 0.269*** | | |
| | (39.01) | (43.99) | | | (240.41) | (615.51) | | |
| *Controls* | Yes | Yes | Yes | Yes | Yes | Yes | Yes | Yes |
| Stock *i* ×Quarter FE | Yes | Yes | Yes | Yes | Yes | Yes | Yes | Yes |
| Stock *j* ×Quarter FE | Yes | Yes | Yes | Yes | Yes | Yes | Yes | Yes |
| Observations | 179,505,839 | 179,505,839 | 179,505,839 | 179,505,839 | 139,293,862 | 139,293,862 | 139,293,862 | 139,293,862 |
| Adjusted R2 | 0.034 | 0.708 | 0.033 | 0.034 | 0.396 | 0.528 | 0.393 | 0.393 |

This table presents the results of robustness checks using alternative measures of excess comovement and institution-driven information diffusion. The definitions of the variables are consistent with those in Equation (9). The individual and time fixed effects are included in the regression model. Standard errors are clustered at the stock-pair level. The t-statistics are reported in parentheses. ***, **, and * denote statistical significance at the 1%, 5%, and 10% levels, respectively.

**Table 7**
**Robustness to alternative specifications.**

| | Panel A: Chinese stock market | | | | Panel B: U.S. stock market | | | |
|---|---|---|---|---|---|---|---|---|
| | **Sub-sample1** | **High FE** | **New CV** | **Sub-sample2** | **Sub-sample1** | **High FE** | **New CV** | **Sub-sample2** |
| | (1) | (2) | (3) | (4) | (1) | (2) | (3) | (4) |
| | $Cor^{FF5}$ | $Cor^{FF5}$ | $Cor^{FF5}$ | $Cor^{FF5}$ | $Cor^{FF5}$ | $Cor^{FF5}$ | $Cor^{FF5}$ | $Cor^{FF5}$ |
| *Flow* | 3.373*** | 3.428*** | 2.157*** | 2.981*** | 0.009*** | 0.010*** | 0.005*** | 0.010*** |
| | (55.13) | (56.92) | (39.42) | (73.42) | (22.17) | (23.65) | (13.35) | (23.30) |
| *Inst* | 0.348*** | 0.231*** | 0.163*** | 0.528*** | 0.233*** | 0.241*** | 0.125*** | 0.224*** |
| | (21.99) | (23.41) | (16.35) | (21.27) | (222.74) | (238.11) | (124.80) | (100.14) |
| *Controls* | Yes | Yes | Yes | Yes | Yes | Yes | Yes | Yes |
| Stock *i* ×Quarter FE | Yes | Yes | Yes | Yes | Yes | Yes | Yes | Yes |
| Stock *j* ×Quarter FE | Yes | Yes | Yes | Yes | Yes | Yes | Yes | Yes |
| Observations | 99,759,402 | 179,502,169 | 179,505,839 | 35,992,569 | 100,263,627 | 139,293,862 | 139,293,862 | 21,818,780 |
| Adjusted R2 | 0.047 | 0.416 | 0.052 | 0.048 | 0.396 | 0.403 | 0.424 | 0.407 |

This table presents the results of additional robustness checks examining the relationship between information diffusion and excess comovement. The definitions of the variables are consistent with those in Equation (9). The individual and time fixed effects are included in the regression model. Standard errors are clustered at the stock-pair level. The t-statistics are reported in parentheses. ***, **, and * denote statistical significance at the 1%, 5%, and 10% levels, respectively.

**Table 8**
**Results of Fama-Macbeth Regressions (Newey-West Adjusted).**

| **Panel A: Chinese stock market** | | | | | | |
|---|---|---|---|---|---|---|
| Variables | *Flow* | *Inst* | *News* | *Analyst* | *Similarroa* | *Similarlever* |
| Coefficient | 129.334*** | 2.482*** | 0.003*** | 0.032*** | 0.003*** | 0.006*** |
| T-statistics | 35.0737 | 3.928 | 14.045 | 14.653 | 16.529 | 4.725 |
| Variables | *Similarsize* | *Similarprice* | *Similarmom* | *Similarvol* | *Samelocation* | *Sameind* |
| Coefficient | 0.003*** | 0.012*** | 0.012*** | 0.006*** | 0.006*** | 0.007*** |
| T-statistics | 6.045 | 8.904 | 11.283 | 5.723 | 6.048 | 7.160 |
| **Panel B: U.S. stock market** | | | | | | |
| Variables | *Flow* | *Inst* | *News* | *Analyst* | *Similarroa* | *Similarlever* |
| Coefficient | 0.276* | 0.260*** | 0.016*** | 0.026*** | 0.005*** | 0.002*** |
| T-statistics | 1.919 | 7.303 | 3.931 | 20.694 | 10.688 | 8.172 |
| Variables | *Similarsize* | *Similarprice* | *Similarmom* | *Similarvol* | *Samelocation* | *Sameind* |
| Coefficient | 0.012*** | 0.008*** | 0.006*** | 0.005*** | 0.005*** | 0.034*** |
| T-statistics | 21.929 | 19.651 | 13.333 | 10.019 | 3.095 | 23.527 |

This table reports Fama-Macbeth regression coefficients examining the relationship between investor-driven information diffusion and excess comovement. The definitions of the variables are consistent with those in Equation (9). The reported coefficients represent time-series averages of quarterly estimates, with Newey-West adjusted t-statistics (lag=3) accounting for autocorrelation in parentheses. ***, **, and * denote statistical significance at the 1%, 5%, and 10% levels, respectively.

**Table 9**
**Results of mechanism analysis.**

**Panel A: Chinese stock market**

| | (1) | (2) | (3) | (4) | (5) |
|---|---|---|---|---|---|
| | $Cor^{FF5}$ | $COR_{BSI}^{Retl}$ | $Cor^{FF5}$ | $COR_{BSI}^{Inst}$ | $Cor^{FF5}$ |
| $Flow$ | 3.577*** | 2.951*** | 2.469*** | 2.899*** | 2.741*** |
| | (81.90) | (82.97) | (79.02) | (77.92) | (73.43) |
| $Inst$ | 0.776*** | 1.235*** | 0.381*** | 1.629*** | 0.307*** |
| | (36.46) | (42.10) | (28.83) | (40.56) | (22.12) |
| $COR_{BSI}^{Retl}$ | | | 0.379*** | | |
| | | | (4,016.61) | | |
| $COR_{BSI}^{Inst}$ | | | | | 0.311*** |
| | | | | | (3,091.37) |
| *Controls* | Yes | Yes | Yes | Yes | Yes |
| Stock $i$×Quarter FE | Yes | Yes | Yes | Yes | Yes |
| Stock $j$×Quarter FE | Yes | Yes | Yes | Yes | Yes |
| Observations | 179,505,839 | 179,505,839 | 179,505,839 | 179,505,839 | 179,505,839 |
| Adjusted R2 | 0.033 | 0.342 | 0.141 | 0.277 | 0.106 |

**Panel B: U.S. stock market**

| | (1) | (2) | (3) | (4) | (5) |
|---|---|---|---|---|---|
| | $Cor^{FF5}$ | $COR_{BSI}^{Retl}$ | $Cor^{FF5}$ | $COR_{BSI}^{Inst}$ | $Cor^{FF5}$ |
| $Flow$ | 0.009*** | 0.007*** | 0.008*** | 0.003*** | 0.008*** |
| | (23.23) | (28.64) | (33.68) | (14.59) | (33.38) |
| $Inst$ | 0.243*** | 0.017*** | 0.243*** | 0.054*** | 0.119*** |
| | (239.88) | (38.77) | (239.53) | (56.40) | (223.49) |
| $COR_{BSI}^{Retl}$ | | | 0.005*** | | |
| | | | (49.03) | | |
| $COR_{BSI}^{Inst}$ | | | | | 0.002*** |
| | | | | | (33.41) |
| *Controls* | Yes | Yes | Yes | Yes | Yes |
| Stock $i$×Quarter FE | Yes | Yes | Yes | Yes | Yes |
| Stock $j$×Quarter FE | Yes | Yes | Yes | Yes | Yes |
| Observations | 139,293,862 | 139,293,862 | 139,293,862 | 139,293,862 | 139,293,862 |
| Adjusted R2 | 0.393 | 0.013 | 0.235 | 0.029 | 0.269 |

This table reports the results of mechanism analysis. $COR_{BSI}^{Retl}$ and $COR_{BSI}^{Inst}$ are the mediator variables for retail and institutional investors. The definitions of the variables are consistent with those in Equation (9). The individual and time fixed effects are included in the regression model. Standard errors are clustered at the stock-pair level. The t-statistics are reported in parentheses. ***, **, and * denote statistical significance at the 1%, 5%, and 10% levels, respectively.

**Table 10**

**Comparison of excess comovement across information diffusion speed portfolios.**

| Panel A: Chinese stock market | | | Panel B: U.S. stock market | | |
|---|---|---|---|---|---|
| | Mean | Median | | Mean | Median |
| ***fast*** | 0.019216 | 0.018121 | ***fast*** | 0.177183 | 0.149316 |
| ***2*** | 0.006094 | 0.005339 | ***2*** | 0.187276 | 0.157461 |
| ***3*** | 0.004914 | 0.003485 | ***3*** | 0.184604 | 0.149073 |
| ***4*** | 0.004594 | 0.001889 | ***4*** | 0.157314 | 0.125916 |
| ***slow*** | 0.005575 | 0.001519 | ***slow*** | 0.140972 | 0.104311 |
| ***fast–slow*** | 0.013641*** | 0.016601*** | ***fast–slow*** | 0.036211* | 0.045006* |

This table reports the mean and median of each group's weighted average correlation coefficient for the Chinese and U.S. stock markets. We sort stock pairs into quintiles based on the level of monthly retail investors' information diffusion. *Slow* denotes the coefficient for the group with the slowest information diffusion, while *fast* represents the coefficient for the group with the fastest information diffusion. The variable *fast–slow* denotes the result of the non-parametric test comparing the fastest and slowest group. ***, **, and * denote statistical significance at the 1%, 5%, and 10% levels, respectively.

**Table 11**

**Results of the lead-lag regression analysis.**

| | ***Constant*** | $COR_{t-1}^{fast}$ | $R^2$ |
|---|---|---|---|
| Chinese stock market | 0.003** | 0.291*** | 0.105 |
| U.S. stock market | 0.024** | 0.082* | 0.348 |

This table reports the results of testing the lead-lag relationship in excess comovement, as specified in Equation (10). The dependent variable is the contemporaneous correlation of the slow-diffusion group ($COR_t^{slow}$). The key independent variable is the one-month-lagged correlation of the fast-diffusion group ($COR_{t-1}^{fast}$). ***, **, and * indicate that the coefficients are significant at the 1%, 5%, and 10% levels, respectively.

**Table 12**
**Results on the predictability of information diffusion for excess comovement.**

| | Panel A: Chinese stock market | | | | Panel B: U.S. stock market | | | |
|---|---|---|---|---|---|---|---|---|
| | (1) | (2) | (3) | (4) | (5) | (6) | (7) | (8) |
| | $Cor^{FF5}$ | $Cor^{FF5}$ | $Cor^{FF5}$ | $Cor^{FF5}$ | $Cor^{FF5}$ | $Cor^{FF5}$ | $Cor^{FF5}$ | $Cor^{FF5}$ |
| *L. Flow* | 3.459*** | 2.544*** | 2.451*** | 2.349*** | 0.012*** | 0.002*** | 0.003*** | 0.004*** |
| | (77.81) | (52.29) | (50.71) | (50.51) | (27.02) | (3.45) | (3.86) | (5.27) |
| *L2. Flow* | | 1.419*** | 0.637*** | 0.571*** | | 0.001 | 0.0007 | 0.0004 |
| | | (30.75) | (12.91) | (12.06) | | (1.09) | (0.74) | (0.44) |
| *L3. Flow* | | | 1.420*** | 0.753*** | | | 0.0003 | 0.0002 |
| | | | (32.28) | (16.66) | | | (0.65) | (0.22) |
| *L4. Flow* | | | | 1.321*** | | | | 0.009 |
| | | | | (34.48) | | | | (1.28) |
| *L. Inst* | 0.259*** | 0.179*** | 0.279*** | 0.456*** | 0.191*** | 0.096*** | 0.099*** | 0.106*** |
| | (16.54) | (2.61) | (4.14) | (5.19) | (185.02) | (43.99) | (43.30) | (43.92) |
| *L2. Inst* | | 0.134** | 0.142 | 0.094 | | 0.109*** | 0.040*** | 0.039*** |
| | | (2.06) | (1.26) | (0.75) | | (52.96) | (13.46) | (12.53) |
| *L3. Inst* | | | 0.009 | 0.009 | | | 0.073*** | 0.044*** |
| | | | (0.11) | (0.07) | | | (31.48) | (14.41) |
| *L4. Inst* | | | | 0.1465 | | | | 0.027*** |
| | | | | (1.64) | | | | (13.35) |
| *Controls* | Yes | Yes | Yes | Yes | Yes | Yes | Yes | Yes |
| Stock *i* ×Quarter FE | Yes | Yes | Yes | Yes | Yes | Yes | Yes | Yes |
| Stock *j* ×Quarter FE | Yes | Yes | Yes | Yes | Yes | Yes | Yes | Yes |
| Observations | 165,337,435 | 153,097,589 | 140,467,517 | 128,171,735 | 96,013,039 | 88,626,857 | 81,746,343 | 75,453,431 |
| Adjusted R2 | 0.034 | 0.034 | 0.034 | 0.035 | 0.400 | 0.401 | 0.403 | 0.404 |

This table reports the regression results examining the predictability of information diffusion for excess comovement in the Chinese and U.S. stock markets, respectively. The definitions of the variables are consistent with those in Equation (9). The individual and time fixed effects are included in the regression model. Standard errors are clustered at the stock-pair level. The t-statistics are reported in parentheses. ***, **, and * denote statistical significance at the 1%, 5%, and 10% levels, respectively.

## Appendix A: Variable definition

| Variables | Definition |
|---|---|
| **Dependent variables** | |
| $Cor^{FF5}$ | The Pearson correlation coefficient of the daily residual returns from the Fama-French five-factor (FF5) model for stocks i and stock j during the quarter. |
| $Cor^{FF3}$ | The Pearson correlation coefficient of the daily residual returns from the Fama-French three-factor (FF3) model for stocks i and stock j during the quarter. |
| $Cor^{CAPM}$ | The Pearson correlation coefficient of the daily residual returns from the Capital Asset Pricing Model (CAPM) for stocks i and stock j during the quarter. |
| **Independent variables** | |
| *Flow* | The number of co-investors' posts and replies, as well as the subsequent replies triggered by those posts and replies for stock i and stock j during the quarter. |
| *Inst* | The sum of ownership proportions across all institutions that hold both stocks i and stock j during the quarter. |
| *Inst* (1%) | The sum of ownership proportions across all institutions that hold more than 1% of both stocks i and stock j during the quarter. |
| *Inst* (3%) | The sum of ownership proportions across all institutions that hold more than 3% of both stocks i and stock j during the quarter. |
| **Mediating variables** | |
| $COR_{BSI}^{Inst}$ | The Pearson correlation coefficient of the daily BSI index of institutional investors for stocks i and stock j during the quarter. |
| $COR_{BSI}^{Retl}$ | The Pearson correlation coefficient of the daily BSI index of retail investors for stocks i and stock j during the quarter. |
| **Instrumental variable** | |
| *Age* | The average registration length of all co-investors for stocks i and stocks j during the quarter. |
| *PostUpdate* | An indicator variable that takes the value of 1 from the fourth quarter of 2013 onward, and 0 otherwise. |
| *Dmerger* | An indicator variable that takes the value of 1 in the quarters following an institutional merger event, and 0 otherwise. |
| **Control variables** | |
| *News* | The number of news articles that mention both stock i and stock j during the quarter. |
| *Analyst* | The number of analysts covering both stock i and stock j during the quarter. |
| *Similarind* | An indicator variable taking the value of 1 if stock i and stock j share the same industry classification during the quarter and 0 otherwise. |
| *Similarlocation* | An indicator variable taking the value of 1 if the headquarter of stock i and stock j are located in the same city during the quarter and 0 otherwise. |
| *Similarprice* | An indicator variable taking the value of 1 if stock i and stock j share the same decile of stock price during the quarter and 0 otherwise. |
| *Similarsize* | An indicator variable taking the value of 1 if stock i and stock j share the same decile of market value of equity during the quarter and 0 otherwise. |
| *Similarroa* | An indicator variable taking the value of 1 if stock i and stock j share the same decile of net profit scaled by total assets during the quarter and 0 otherwise. |
| *Similarlever* | An indicator variable taking the value of 1 if stock i and stock j share the same decile of total debts scaled by total assets during the quarter and 0 otherwise. |
| *Similarvol* | An indicator variable taking the value of 1 if stock i and stock j share the same decile of the standard deviation of daily returns during the quarter and 0 otherwise. |
| *Similarmom* | An indicator variable taking the value of 1 if stock i and stock j share the same |

| | decile of annual stock return during the quarter and 0 otherwise. |
|---|---|
| *Diffprice* | The natural logarithm of the absolute difference in stock price between stock i and stock j during the quarter. |
| *Diffsize* | The natural logarithm of the absolute difference in the market value of equity between stock i and stock j during the quarter. |
| *Diffroa* | The natural logarithm of the absolute difference in net profit scaled by total assets between stock i and stock j during the quarter. |
| *Difflever* | The natural logarithm of the absolute difference in total debts scaled by total assets between stock i and stock j during the quarter. |
| *Diffvol* | The natural logarithm of the absolute difference in the standard deviation of daily returns within the quarter between stock i and stock j. |
| *Diffmom* | The natural logarithm of the absolute difference in annual stock return between stock i and stock j during the quarter. |